\documentclass[12pt,a4paper,twoside]{article}
\usepackage{amsmath}
\usepackage{amssymb}
\usepackage{fancyhdr}
\usepackage{setspace}
\usepackage[final]{showkeys}
\usepackage[dvips]{graphicx}
\usepackage{bbm}

\newcommand{\Cx}{\mathbbm{C}}
\newcommand{\Ir}{\mathbbm{Z}}
\newcommand{\Nl}{\mathbbm{N}}
\newcommand{\Rl}{{\mathbbm R}}
\newcommand{\Ts}{\mathbbm{T}}
\newcommand{\idty}{\mathbbm{1}}

\newcommand{\coh}[2]{\ensuremath{|C_{#1}(#2) \rangle}}
\newcommand{\lcoh}[2]{\ensuremath{\langle C_{#1}(#2)|}}
\newcommand{\<}{\langle}
\renewcommand{\>}{\rangle}
\newcommand{\bi}[1]{\boldsymbol{#1}}
\renewcommand{\c}[1]{\mathcal{#1}}
\newcommand{\g}[1]{\mathfrak{#1}}
\renewcommand{\r}[1]{\mathrm{#1}}
\newcommand{\co}[1]{\textsf{#1}}
\DeclareMathOperator*{\tr}{Tr}

\newenvironment{proof}{\setlength{\parindent}{0pt}{\bf Proof:} \par }{\par %
  \hfill $\blacksquare$ \par}
\newtheorem{definition}{Definition}[section]
\newtheorem{proposition}{Proposition}[section]
\newtheorem{condition}{Condition}[section]
\newtheorem{lemma}{Lemma}[section]
\newtheorem{theorem}{Theorem}[section]%

\begin{document}
\ \vskip 1cm

\baselineskip = 22pt
\pagestyle{fancy}
\renewcommand{\headrulewidth}[0]{0pt}
\lhead[\fancyplain{}{\footnotesize \thepage \hspace{5mm}{\it
F.~Benatti, V.~Cappellini, M.~De~Cock, M.~Fannes and D.~Vanpeteghem}}]%
      {\fancyplain{}{}}
\rhead[\fancyplain{}{}]%
      {\fancyplain{}{\footnotesize {\it Classical Limit of Quantum Dynamical Entropies}\hspace{5mm}\thepage}}
\chead{}\lfoot{}\cfoot{}\rfoot{}
\onehalfspacing
\centerline{\LARGE \bf Classical Limit of Quantum Dynamical Entropies} 
\bigskip

\centerline{\large F.~Benatti$^1$, V.~Cappellini$^1$,
M.~De~Cock$^{\,2}$, M.~Fannes$^{\,2}$ and 
D.~Vanpeteghem$^{2,\,}$\footnote[3]{Research Assistant of the Fund for 
Scientific Research - Flanders (Belgium)(F.W.O. - Vlaanderen)}}
\bigskip

\centerline{$^1$\,Dip. Fisica Teorica}
\centerline{Universit\`a di Trieste, Strada Costiera 11, I-34014 Trieste, Italy}
\bigskip

\centerline{$^2$\,Instituut voor Theoretische Fysica}
\centerline{K.U. Leuven, B-3001 Leuven, Belgium}
\bigskip\bigskip

\noindent
\textbf{Abstract}
Two non-commutative dynamical entropies are studied in connection
with the classical limit.  For systems with a strongly chaotic
classical limit, the Kolmogorov-Sinai invariant is recovered on time
scales that are logarithmic in the quantization parameter. The model
of the quantized hyperbolic automorphisms of the 2-torus is examined
in detail.
\bigskip


\section{Introduction}   
\label{s1}

Classical chaos is understood as motion on compact regions with
trajectories highly sensitive to initial
conditions~\cite{Sch95:1,Gia89:1,Cas95:1,Zas85:1}. Once quantized, the motion has
discrete energy spectrum and  behaves almost periodically in time.
Nevertheless, nature is fundamentally quantal and, according to the
correspondence principle, classical behaviour emerges in the limit
$\hbar\to0$.

Also, classical and quantum mechanics are expected to almost coincide
in the semi-classical regime, that is over times scaling as
$\hbar^{-\alpha}$ for some $\alpha>0$~\cite{Zas85:1}. Actually, this is
true only for regular classical limits,  while for chaotic ones the
semi-classical regime typically scales as 
$-\!\log\hbar$~\cite{Gia89:1,Cas95:1,Zas85:1}. Both time scales diverge  when
$\hbar\to0$, but the shortness of the latter means that classical
mechanics has to be replaced by quantum mechanics much sooner for
quantum systems with chaotic classical behaviour. The logarithmic
breaking time $-\!\log{\hbar}$ has been considered  by some as a
violation of the correspondence principle~\cite{For91:1,For92:2}, by others,
see~\cite{Cas95:1} and Chirikov in~\cite{Gia89:1}, as the evidence that time
and classical limits do not commute.

The analytic studies of logarithmic time scales have been mainly
performed  by means of semi-classical tools, essentially by focusing,
via coherent state techniques,  on the phase space localization  of
specific time evolving quantum observables. In the following, we
shall show how they emerge in the context of quantum dynamical
entropies. As a particular example, we shall concentrate on finite
dimensional quantizations of hyperbolic automorphisms of the 2-torus,
which are prototypes of chaotic behaviour; indeed, their trajectories
separate exponentially fast with a Lyapounov exponent
$\log\lambda>0$~\cite{Arn68:1,Wal82:1}. Standard quantization, \`a la Berry,
of hyperbolic automorphisms~\cite{Ber79:1,Deg93:1} yields Hilbert spaces of a
finite dimension~$N$. This dimension plays the role of semi-classical
parameter and sets the minimal size $1/N$ of quantum phase space
cells. 

By the theorems of Ruelle and Pesin~\cite{Man87:1}, the positive Lyapounov
exponents of smooth, classical dynamical systems  are related to the
dynamical entropy of Kolmogorov~\cite{Kat99:1} which measures the
information per time step provided by the dynamics. There are several
candidates for non-commutative extensions of the
latter~\cite{Con87:1,Ali94:1,Voi92:1,Acc97:1,Slo94:1}: in  this paper we shall
use two of 
them~\cite{Con87:1,Ali94:1}  and study their semi-classical limit.   We show
that, from both of them, one recovers the Kolmogorov-Sinai entropy
by  computing the average quantum entropy produced over a logarithmic
time scale and then taking the classical limit. This confirms the
numerical results in~\cite{Ali96:2}, where the dynamical
entropy~\cite{Ali94:1} is applied to the study of  the quantum kicked
top. In this approach, the presence of logarithmic time scales
indicates the  typical scaling for a joint time--classical limit
suited to preserve positive  entropy production in quantized
classically chaotic quantum systems.

The paper is organized as follows: Section~\ref{s2} contains a brief
review of the algebraic approach to classical and dynamical systems,
while Section~\ref{s3} introduces some basic semi-classical tools.
Sections~\ref{s4} and~\ref{s5} deal with the quantization of
hyperbolic maps on finite  dimensional Hilbert spaces and the
relation between classical and time limits. Section~\ref{s6} gives an
overview of the quantum dynamical entropy of Connes, Narnhofer and
Thirring~\cite{Con87:1}(CNT--entropy) and of Alicki and
Fannes~\cite{Ali01:1,Ali94:1} (ALF-entropy, where L stands for Lindblad); 
finally, in Section~\ref{s7} their semi-classical behaviour is
studied and  the emergence of a typical logarithmic time scale is
showed.

\section{Dynamical systems: algebraic setting}
\label{s2}

We consider reversible, discrete time, compact classical dynamical
systems that can be represented by a triple $(\c X,T,\mu)$, where:
\begin{itemize}
\item
 $\c X$ is a compact metric space: the phase space of the system.
\item
 $T$ is a measurable transformation of $\c X$ that is invertible such
 that $T^{-1}$ is also measurable. The group $\{T^k \mid k\in\Ir\}$ implements 
 the conservative dynamics in discrete time.
\item
 $\mu$ is a $T$-invariant probability measure on $\c X$, i.e.\ 
 $\mu\circ T = \mu$. 
\end{itemize}

In this paper, we consider a general scheme for quantizing and 
dequantizing, i.e.\ for taking the classical limit (see~\cite{Wer95:1}). 
Within this framework, we focus on the semi-classical
limit of quantum dynamical entropies of 
finite dimensional quantizations of the Arnold cat map and of generic
hyperbolic automorphisms of the 2-torus, cat maps
for short.
In order to make the quantization procedure more explicit, it proves useful
to follow an algebraic approach and replace $(\c X,T,\mu)$ with
$(\g M_\mu,\Theta,\omega_\mu)$ where
\begin{itemize}
\item
 $\g M_\mu$ is the von~Neumann algebra $\g L_ \mu^\infty(\c X)$ of
 (equivalence classes of) essentially bounded $\mu$-measurable
 functions on $\c X$, equipped with the so-called essential supremum
 norm $\|\cdot\|_\infty$~\cite{Rud87:1}.
\item
 $\omega_\mu$ is the state on $\g M_\mu$ defined by the reference
 measure $\mu$ 
 \begin{equation*}
  \omega_\mu(f):= \int_{\c X}\mu(dx)\, f(x).
 \end{equation*} 
\item
 $\{\Theta^k \mid k\in\Ir\}$ is the discrete group of automorphisms
 of $\g M_\mu$ which implements the dynamics: $\Theta(f) := f \circ
 T^{-1}$. The invariance of the reference measure reads now 
 $\omega_\mu \circ \Theta = \omega_\mu$.
\end{itemize}

Quantum dynamical systems are described in a completely similar way
by a triple $(\g M,\Theta,\omega)$, the critical difference being
that the algebra of observables $\g M$ is no longer Abelian:
\begin{itemize}
\item
 $\g M$ is a von~Neumann algebra of operators, the observables, acting on 
 a Hilbert space $\g H$.
\item
 $\Theta$ is an automorphism of $\g M$.
\item
 $\omega$ is an invariant normal state on $\g M$: $\omega\circ\Theta=\omega$.
\end{itemize}

Quantizing essentially corresponds to suitably mapping the commutative,
classical triple $(\g M_\mu,\Theta,\omega_\mu)$ to a non-commutative,
quantum triple $(\g M,\Theta,\omega)$. 

\section{Classical limit: coherent states}
\label{s3}

Performing the classical limit or a semi-classical analysis consists 
in studying how a family of
algebraic triples $(\g M,\Theta,\omega)$ 
depending on a quantization $\hbar$-like parameter  
is mapped onto $(\g M_\mu,\Theta,\omega_\mu)$
when the parameter goes to zero. 
The most successful semi-classical tools are based on
the use of coherent states. 

For our purposes, we shall use a large integer $N$ as a quantization parameter, i.e.\ we use $1/N$ as the
$\hbar$-like parameter. 
In fact, we shall consider cases where $\g M$ is the
algebra $\c M_N$ of $N$-dimensional square matrices acting on $\Cx^N$,
the quantum reference state is the normalized trace
$\frac{1}{N} \tr\ $ on $\c M_N$,
denoted by $\tau_N$ and the dynamics is given in terms of a unitary
operator $U_T$ on $\Cx^N$ in the standard way: $\Theta_N(X) :=
U_T^*X\,U_T$. 

In full generality, coherent states  will be identified as follows.

\begin{definition}
\label{coh}
 A family $\{\vert C_N(x)\rangle \mid x\in\c X\}\in\g H$ of vectors, 
 indexed by points
 $x\in\c X$, constitutes a set of coherent states if it satisfies the
 following requirements
 \begin{enumerate}
 \item
  \co{Measurability}: $x \mapsto \vert C_N(x)\rangle$ is measurable on $\c X$;
 \item
  \co{Normalization}: $\|C_N(x)\|^2 = 1$, $x\in\c X$;
 \item
  \co{Overcompleteness}: $N \int_{\c X}\mu(dx)\, \coh{N}{x}
  \lcoh{N}{x} = \idty$;
 \item
  \co{Localization}: given $\varepsilon>0$ and $d_0>0$, there exists 
  $N_0(\epsilon,d_0)$ such that for $N\ge N_0$ and $d(x,y)\ge d_0$ one has
  \begin{equation*} 
   N |\< C_N(x), C_N(y) \>|^2 \le \varepsilon.
  \end{equation*}
 \end{enumerate}
\end{definition}

The overcompleteness condition may be written in dual form as
\begin{equation*}
 N \int_{\c X}\mu(dx)\, \<C_N(x), X\, C_N(x)\> = \tr X, \quad X\in\c
 M_N.
\end{equation*}
Indeed,
\begin{equation*}
 N \int_{\c X}\mu(dx)\, \<C_N(x), X\, C_N(x)\> = N \tr \int_{\c
 X}\mu(dx)\, \coh{N}{x} \lcoh{N}{x}\, X = \tr X. 
\end{equation*}

\subsection{Anti-Wick Quantization}

In order to study the classical limit and, more generally,
semi-classical behaviour of $(\c M_N ,\Theta_N,\tau_N)$
when $N\to\infty$, we introduce two linear maps. The first,
$\gamma_{N\infty}$, (anti-Wick quantization) associates $N\times N$
matrices to functions in $\g M_\mu{\bf =\g L_ \mu^\infty(\c X)}$, 
the second one, $\gamma_{\infty N}$, maps $N\times N$ matrices to functions 
in $\g L_ \mu^\infty(\c X)$.

\begin{definition}
\label{qWick}
 Given a family $\{\vert C_N(x)\rangle \mid x\in\c X\}$ of coherent states in
 $\Cx^N$, the anti-Wick quantization scheme will be described by a
 (completely) positive unital map $\gamma_{N\infty}: \g M_\mu\to\c
 M_N$ 
 \begin{equation*}
 {\bf \g M_\mu\ni} f \mapsto   
 N \int_{\c X}\mu(dx)\, f(x)\,
  \coh{N}{x} \lcoh{N}{x}=:\gamma_{N\infty}(f)\in \c M_N\quad .
 \end{equation*}
 The corresponding dequantizing map $\gamma_{\infty N}: \c M_N\to\g M_\mu$
 will correspond to the (completely) positive unital map
 \begin{equation*}
  \c M_N \ni X \mapsto 
  \<C_N(x), X\,C_N(x)\>=:\gamma_{\infty N}(X)(x) \in \g M_\mu
  \quad .
 \end{equation*}
\end{definition}

Both maps are identity preserving because of the conditions imposed
on the family of coherent states and are also completely positive
since the domain of $\gamma_{N\infty}$ is a commutative algebra as
well as the range of $\gamma_{\infty N}$. Moreover, 
\begin{equation}
 \bigl\| \gamma_{\infty N} \circ \gamma_{N\infty}(g) \bigr\|_\infty \le
 \|g\|_\infty,\quad g\in\g M_\mu\quad ,
\label{bound}
\end{equation}
where $\|\cdot\|_\infty$ denotes the essential norm on
$\g M_\mu=\g L_\mu^\infty(\c X)$.
The following two equivalent properties are less trivial:

\begin{proposition}
\label{prop1}
 For all $f\in\g M_\mu$
 \begin{equation*}
  \lim_{N\to\infty} \gamma_{\infty N} \circ \gamma_{N\infty}(f) =
  f\quad \mu\text{-a.e.}
 \end{equation*}
\end{proposition}
\medskip

\begin{proposition}
\label{prop2}
 For all $f,g\in\g M_\mu$
 \begin{equation*}
  \lim_{N\to\infty} \tau_N \bigl( \gamma_{N\infty}(f)^*
  \gamma_{N\infty}(g) \bigr) = \omega_\mu(\overline f g) = \int_{\c X}
  \mu(dx)\, \overline{f(x)}g(x).
 \end{equation*}
\end{proposition}
\medskip

\noindent
The previous two propositions can be taken as requests on any
well--defined quantization--dequantization scheme for observables.
In the sequel, we shall need the notion of quantum dynamical systems
$(\c M_N,\Theta_N,\tau_N)$ tending to the classical limit $(\c
X,T,\mu)$. We then not only need convergence of observables but also
of the dynamics. This aspect will be considered in Section~\ref{s5}. 

{\setlength{\parindent}{0pt}  
\newpage
\textbf{Proof of Proposition~\ref{prop1}:}

We first prove the assertion when $f$ is continuous on $\c X$ and
then remove this condition. We show that the quantity
\begin{align*}
 F_N(x)
 &:= \Bigl|f(x) - \gamma_{\infty N} \circ \gamma_{N\infty}(f)(x) \Bigr|
 \\
 &= \left|f(x) - N \int_{\c X}\mu(dy)\, f(y)\, |\< C_N(x), C_N(y)\>|^2
 \right| \\
 &= N\left| \int_{\c X }\mu(dy)\, (f(y) - f(x))\, |\< C_N(x), C_N(y)\>|^2 
 \right|
\end{align*}
becomes arbitrarily small for $N$ large enough, uniformly in $x$.
Selecting a ball $B(x,d_0)$ of radius $d_0$, using the mean-value
theorem and property~(\ref{coh}.3), we derive the upper bound
\begin{align}
 F_N(x)
 &\le N\left|\int_{B(x,d_0)} \mu(dy)\, (f(y) - f(x))\, |\< C_N(x),
 C_N(y)\>|^2 \right|
\nonumber \\
 &\quad + N\left|\int_{\c X\setminus B(x,d_0)} \mu(dy)\, (f(y) -
 f(x))\, |\< C_N(x), C_N(y)\>^2 \right| \\
 &\le |f(c) - f(x)|
 + \int_{\c X\setminus B(x,d_0)} \mu(dy)\, |f(y) - f(x)|\, N\,
 |\< C_N(x), C_N(y)\>|^2,
\label{prop1.2}
\end{align}
where $c\in B(x,d_0)$.

Because $\c X$ is compact, $f$ is uniformly continuous. Therefore, we
can choose $d_0$ in such a way that $|f(c) - f(x)| < \varepsilon$
uniformly in $x\in\c X$. On the other hand, from the localization
property~(\ref{coh}.4), given $\varepsilon'>0$, there exists an
integer $N_0(\varepsilon', d_0)$ such that $N|\< C_N(x), C_N(y)\>|^2 
< \varepsilon'$ whenever $N>N_0(\varepsilon',d_0)$. This choice
leads to the upper bound
\begin{align}
 F_N(x)
 &\le \varepsilon + \varepsilon'  \int_{\c X\setminus B(x,d_0)}
 \mu(dy)\, |f(y) - f(x)|
\nonumber\\
 &\le \varepsilon + \varepsilon'  \int_{\c X} \mu(dy)\, |f(y) - f(x)| \le
 \varepsilon + 2 \varepsilon' \|f\|_\infty.
\end{align}

To get rid of the continuity of $f$, we use Lusin's theorem~\cite{Rud87:1}. It
states that, given $f \in \g L_\mu^\infty(\c X)$,
with $\c X$ compact,
there exists a sequence $\{f_n\}$ of continuous functions on $\c X$
such that $|f_n| \le \|f\|_\infty$ and converging to $f$ $\mu$-almost
everywhere. 
Thus, for $f \in \g L_\mu^\infty(\c X)$, we pick such a
sequence and estimate
\begin{align*}
 F_N(x)
 &\le |f(x)-f_n(x)| + \Bigl|f_n(x) - \gamma_{\infty
 N}\circ\gamma_{N\infty}(f_n)(x) \Bigr| \\
 &\quad+ \Bigl|\gamma_{\infty N} \circ \gamma_{N\infty}(f_n- f)(x)\Bigr|.
\end{align*}
The first term can be made arbitrarily small ($\mu.a.e$) by choosing $n$ large
enough because of Lusin's theorem, while the second one goes to $0$
when  $N\to\infty$ since $f_n$ is continuous.
Finally, the third term becomes as well vanishingly small with $n\to\infty$
as one can deduce from
\begin{align*}
 &\int_{\c X} \mu(dx)\,  \Bigl|\gamma_{\infty N} \circ \gamma_{N\infty}
 (f - f_n)(x)\Bigr| \\
 &\quad= \int_{\c X} \mu(dx)\, \left| \int_{\c X} \mu(dy)\, (f(y) -
 f_n(y))\, N\, |\< C_N(x), C_N(y)\>|^2 \right| \\
 &\quad\le \int_{\c X} \mu(dy)\, |f(y) - f_n(y)|\, \int_{\c X}
 \mu(dx)\, N\, |\< C_N(x), C_N(y)\>|^2 \\
 &\quad= \int_{\c X} \mu(dy)\, |f(y) - f_n(y)|,
\end{align*}
where exchange of integration order is harmless
because of the existence of the integral~(\ref{bound}).
The last integral goes to zero with $n$ by dominated convergence and thus
the result follows.
\hfill $\blacksquare$ \par}
\medskip

{\setlength{\parindent}{0pt}  
\textbf{Proof of Proposition~\ref{prop2}:}

Consider
\begin{align*}
 \Omega_N
 &:= \Bigl| \tau_N \bigl(\gamma_{N\infty}(f)^*
 \gamma_{N\infty}(g)\bigr) - \omega_\mu(\overline f g)\Bigr| \\
 &=N \left| \int_{\c X} \mu(dx)\, \overline{f(x)} \int_{\c X}
 \mu(dy)\, (g(y)-g(x))\, |\< C_N(x), C_N(y)\>|^2 \right| \\
 &\le \int_{\c X} \mu(dx)\, |f(x)|\, \left| \int_{\c X} \mu(dy)\,
 (g(y)-g(x))\, N\, |\< C_N(x), C_N(y)\>|^2 \right|.
\end{align*}
By choosing a sequence of continuous $g_n$ approximating {\bf $g\in\g
L_\mu^\infty(\c X)$}, and arguing as in the previous
proof, we get the following upper bound:
\begin{align*}
 \Omega_N
 &\le N \int_{\c X} \mu(dx)\, |f(x)|\,
 \left| \int_{\c X} \mu(dy)\, (g(y)-g_n(y))\, |\< C_N(x), C_N(y)\>|^2
 \right| \\
 &+ N \int_{\c X} \mu(dx)\, |f(x)|\, \left|
 \int_{\c X} \mu(dy)\, (g_n(y)-g_n(x))\, |\< C_N(x), C_N(y)\>|^2
 \right| \\
 &+ N \int_{\c X} \mu(dx)\, |f(x)|\, \left|
 \int_{\c X} \mu(dy)\, (g(x)-g_n(x))\, |\< C_N(x), C_N(y)\>|^2 \right|.
\end{align*}
The integrals in the first and third lines go to zero by dominated 
convergence and Lusin's theorem. As regards the middle line, one can
apply the argument used for the quantity $F_N(x)$ in the proof of
Proposition~\ref{prop1}.

\hfill $\blacksquare$ \par}
\medskip

\section{Classical and quantum cat maps}
\label{s4}

In this section, we collect the basic material needed to describe both
classical and quantum cat maps and we introduce a specific set of
coherent states that will enable us to perform the semi-classical
analysis of the dynamical entropy.

\subsection{Finite dimensional quantizations}

We first introduce cat maps in the spirit of the algebraic formulation 
introduced in the previous sections.

\begin{definition}
 \label{ccat-map}
 Hyperbolic automorphisms of the torus, i.e.\ cat maps, are generically
 represented by triples $(\g M_\mu,\Theta,\omega_\mu)$, where
 \begin{itemize}
 \item
  $\g M_\mu$ is the algebra of
  essentially bounded functions on the two dimensional torus
  $\Ts := \Bigl\{ \bi x=(x_1,x_2)\in\Rl^2\pmod{1} \Bigr\}$, equipped 
  with the Lebesgue measure $\mu(d\bi x) := d\bi x$.
 \item
  $\{\Theta^k \mid k\in\Ir\}$ is the family of automorphisms
  (discrete time evolution) given by
  $\g M_\mu\ni f \mapsto (\Theta^kf)(\bi x) := f(A^{-k}\,\bi x\pmod{1})$,
  where $A=\begin{pmatrix} a & b \\ c & d\end{pmatrix}$ has integer entries 
  such that $ad-cb=1$, $|a+d|>2$ and maps $\Ts$ onto itself.
 \item
  $\omega_\mu$ is the expectation obtained by integration
  with respect to the Lebesgue measure:
  $\g M_\mu\ni f\mapsto\omega_\mu(f):=\int_{\Ts} d\bi x\, f(\bi x)$, that is
  left invariant by $\Theta$.
 \end{itemize}
\end{definition}


The matrix $A$ has irrational eigenvalues $1<\lambda\ ,\
\lambda^{-1}$, therefore distances stretch along the
eigendirection $\bi u$ of $\lambda$, while shrink along $\bi v$, the
eigendirection of $\lambda^{-1}$. Once the folding 
condition is added, the hyperbolic automorphisms of the torus become
prototypes of classical chaos, with positive Lyapounov exponent
$\log\lambda$.

One can quantize the associated algebraic triple $(\g
M_\mu,\Theta,\omega_\mu)$ on either infinite~\cite{Ben91:2} or finite
dimensional Hilbert spaces~\cite{Ber79:1,Deg93:1,Deb98:1}. 

In the following, we shall
focus on the latter. Given an integer $N$, we consider an orthonormal
basis $\vert j\>$ of $\Cx^N$, where the index $j$ runs through $\Ir_{N}$, 
namely $|j+N\> \equiv |j\>, \ j \in \Ir$. By using this basis we define two unitary
matrices $U_N$ and $V_N$ as follows:
\begin{equation}
 U_N\vert j\> := \exp{(\frac{2\pi i}{N}u)}\vert j+1\>,
 \qquad\text{and}\qquad
 V_N\vert j\> := \exp{\Bigl(\frac{2\pi i}{N}(v-j)\Bigr)} \vert j\>.
\label{UV}
\end{equation}
$u,v\in [0,1)$ are parameters labelling the representations and
\begin{equation}
  U_N^N = \r e^{2i\pi u}\, \idty_N,\quad
  V_N^N = \r e^{2i\pi v}\, \idty_N.
\label{FoldUV}
\end{equation}
It turns out that
\begin{equation}
 U_NV_N=\exp{(\frac{2i\pi}{N})}\, V_NU_N.
\label{algUV}
\end{equation}

Introducing Weyl operators labeled by $\bi n = (n_1,n_2)\in\Ir^2$
\begin{equation}
  W_N(\bi n):= \exp{(\frac{i\pi}{N}n_1n_2)}\, V_N^{n_2}U_N^{n_1} =
  W_N(-\bi n)^*
\label{Weyl1}
\end{equation}
it follows that
\begin{align}
 W_N(N\bi n)
 &= \r e^{i\pi(N n_1n_2+ 2n_1 u+ 2n_2 v)}
\label{Weyl2} \\
 W_N(\bi n)W_N(\bi m)
 &= \exp{(\frac{i\pi}{N}\sigma(\bi n,\bi m))}\, W_N(\bi n+\bi m),
\label{Weyl3}
\end{align}
where $\sigma(\bi n,\bi m) := n_1m_2-n_2m_1$.

\begin{definition}
\label{qcat-map-def}
 Quantized cat maps will be identified with algebraic triples
 $(\c M_N ,\Theta_N,\tau_N)$ where
 \begin{itemize}
 \item
  $\c M_N $ is the full $N\times N$ matrix algebra linearly
  spanned by the Weyl operators $W_N(\bi n)$.
 \item
  $\Theta_N: \c M_N\mapsto \c M_N$ is
  the automorphism such that
  \begin{equation}
   W_N(\bi p) \mapsto\Theta_N(W_N(\bi p)):= W_N(A\bi p),\quad
  \bi p\in\Ir^2.
  \label{dyncat}
  \end{equation}
 \end{itemize}
\end{definition}

In the definition of above, we have omitted reference to
the parameters $u,v$ in~(\ref{UV}): they must be chosen such that
 \begin{equation}
  \begin{pmatrix} a & c \\ b & d \end{pmatrix}
  \begin{pmatrix} u \\ v \end{pmatrix} =
  \begin{pmatrix} u \\ v \end{pmatrix} + 
  \frac{N}{2} \begin{pmatrix} ac \\ bd \end{pmatrix}\pmod{1}.
 \label{uv}
 \end{equation}
Then, the folding condition~(\ref{Weyl2}) is compatible with the 
time evolution~\cite{Deg93:1}.
Further, the algebraic 
relations~(\ref{Weyl3}) are also preserved since the symplectic form
remains invariant, i.e.\ $\sigma(A^t\bi n, A^t\bi m) = \sigma(\bi
n,\bi m)$. 

\noindent
Useful relations can be obtained by using  
\begin{equation}
 W_N(\bi n)\, \vert j\> = \exp(\frac{i\pi}{N}(-n_1n_2+ 2n_1 u+ 2n_2
 v))\, \exp(-\frac{2i\pi}{N}jn_2)\, \vert j+n_1\>\ .
\label{Weyl4}
\end{equation}
From~(\ref{Weyl4}) one readily derives
\begin{align}
 &\tau_N(W_N(\bi n)) = \r e^{\frac{i\pi}{N}(-n_1n_2+ 2n_1 u+ 2n_2 v)}\,
 \delta^{(N)}_{\bi n,0},
\label{Weyl5} \\[6pt]
 &\tau_N(W_N(A\bi n)) = \tau_N(W_N(\bi n)), 
\label{Weyl7} \\
 &\frac{1}{N} \sum_{p_1,p_2=0}^{N-1} W_N(-\bi p)\, W_N(\bi n)\,
 W_N(\bi p) = \tr \Bigl(W_N(\bi n)\Bigr)\,\idty_N,
\label{Weyl8} \\
 &\c M_N \ni X = \sum_{p_1,p_2=0}^{N-1}
 \tau_N \Bigl(X\,W_N(-\bi p)\Bigr)\, W_N(\bi p).
\label{Weyl9}
\end{align}
In~(\ref{Weyl5}), we have introduced the periodic Kronecker delta, that is
$\delta^{(N)}_{{\bf n},0}=1$ if and only if ${\bf n}=0\, \mod(N)$.

From equation~(\ref{Weyl3}) one derives
\begin{equation*}
 [W_N(\bi n), W_N(\bi m)] = 2i\sin\bigl( \frac{\pi}{N}\sigma(\bi n,\bi
 m) \bigr)\, W_N(\bi n+\bi m),
\end{equation*}
which suggests that the $\hbar$-like parameter is $1/N$ and that the
classical limit correspond to $N\to\infty$. In the following section,
we set up a coherent state technique suited to study classical cat
maps as limits of quantized cats.

\subsection{Coherent states for cat maps}

We shall construct a family $\{\vert C_N(\bi x)\rangle \mid \bi x\in\Ts\}$ of
coherent states on the 2-torus by means of the discrete Weyl group.
We define
\begin{equation}
 \vert C_N(\bi x)\rangle := W_N([N \bi x])\, \vert C_N\rangle,
\label{coh1}
\end{equation}
where $[N\bi x] = ([Nx_1],[Nx_2])$,
$0\le[Nx_i]\le N-1$ is the largest integer smaller than $Nx_i$ and
the fundamental vector $\vert C_N\rangle$ is chosen to be
\begin{equation}
 \vert C_N\rangle = \sum_{j=0}^{N-1} C_N(j)\vert j\>,\qquad
 C_N(j):= \frac{1}{2^{(N-1)/2}}\sqrt{\binom{N-1}{j}}.
\label{coh2}
\end{equation}
Measurability and normalization are immediate, over-completeness comes
as follows. 
Let $Y$ be the operator in the left hand side of
property~(\ref{coh}.3). If $\tau_N(Y\, W_N(\bi n)) = \tau_N(W_N(\bi n))$ for
all $\bi n = (n_1,n_2)$ with $0\le n_i\leq N-1$, then according
to~(\ref{Weyl9}) applied to $Y$ it follows that $Y=\idty$. This is
indeed the case as, using~(\ref{Weyl2}) and $N$-periodicity,
\begin{align}
 \tau_N(Y\, W_N(\bi n))
 &= \int_{\Ts} d\bi x\, \<C_N(\bi x), W_N(\bi n)\, C_N(\bi x) \>
\nonumber \\
 &= \int_{\Ts} d\bi x\, \exp{\Bigl(\frac{2\pi i}{N}\sigma(\bi
 n,[N\bi x])\Bigr)}\, \< C_N, W_N(\bi n)\, C_N\>
\nonumber \\
 &= \frac{1}{N^2} \sum_{p_1,p_2=0}^{N-1} \exp{\Bigl( \frac{2\pi
 i}{N}\sigma(\bi n,\bi p)\Bigr)}\, \< C_N, W_N(\bi n)\, C_N\>
\nonumber \\
 &= \tau_N (W_N(\bi n)).
\label{coh3}
\end{align}
In the last line we used that when $\bi x$ runs over $\Ts$,
$[Nx_i]$,  $i=1,2$ runs over the set of integers $0,1,\ldots, N-1$.

The proof the localization property~(\ref{coh}.4) requires several steps.
First, we observe that, due to~(\ref{FoldUV}),
\begin{align}
 E(n)
 &:= \Bigl| \< C_N, W_N(\bi n)\, C_N \> \Bigr|
\nonumber \\
 &= \frac{1}{2^{N-1}} \left| \sum_{\ell=0}^{N-n_1-1}
 \exp\Bigl( -\frac{2\pi i}{N}\ell n_2 \Bigr)
 \sqrt{\binom{N-1}{\ell} \binom{N-1}{\ell+n_1}} \right.
\nonumber \\
 &\quad + \left.\sum_{\ell=N-n_1}^{N-1}
 \exp\Bigl( -\frac{2\pi i}{N}\ell n_2 \Bigr)
 \sqrt{\binom{N-1}{\ell} \binom{N-1}{\ell+n_1-N}} \right|
\label{loc2} \\
 &\le \frac{1}{2^{N-1}} \left[ \sum_{\ell=0}^{N-n_1-1}
 \sqrt{\binom{N-1}{\ell} \binom{N-1}{\ell+n_1}} \right.
\nonumber \\
 &\quad + \left. \sum_{\ell=N-n_1}^{N-1} 
 \sqrt{\binom{N-1}{\ell} \binom{N-1}{\ell+n_1-N}} \right].
\label{loc3}
\end{align}
Second, using the entropic bound of the binomial coefficients
\begin{equation}
 \binom{N-1}{\ell} \le  2^{(N-1) \eta(\frac{\ell}{N-1})}\quad ,
\label{loc4}
\end{equation}
where
\begin{equation}
 \eta(t) := \begin{cases} -t\log_2 t-(1-t)\log_2(1-t) & \text{if } 0<t\leq 1 \\
 0 & \text{if } t=0 \end{cases}\quad ,
\label{loc5}
\end{equation}
we estimate
\begin{align}
 &E(n) \le \frac{1}{2^{N-1}} \left[ \sum_{\ell=0}^{N-1-n_1}
 2^{\frac{N-1}{2} \bigl[ \eta\left(\frac{\ell}{N-1}\right) +
 \eta\left(\frac{\ell+n_1}{N-1}\right) \bigr]} \right.
\nonumber \\
 &\quad+ \left. \sum_{\ell=N-n_1}^{N-1} 2^{\frac{N-1}{2}\bigl[
 \eta\left(\frac{\ell}{N-1}\right) +
 \eta\left(\frac{\ell+n_1-N}{N-1}\right) \bigr]} \right]\quad .
\label{loc6}
\end{align}

The exponents in the two sums are bounded by their maxima
\begin{align}
 \eta\left(\frac{\ell}{N-1}\right) +
 \eta\left(\frac{\ell+n_1}{N-1}\right) 
 &\le 2\eta_1(n_1),\quad \left(0\le\ell\le N-n_1-1\right)
\label{loc6a} \\
 \eta\left(\frac{\ell}{N-1}\right) +
 \eta\left(\frac{\ell+n_1-N}{N-1}\right) 
 &\le 2\eta_2(n_1),\quad \left(N-n_1\le\ell\le N-1\right)
\label{loc6b}
\end{align}
where
\begin{align}
 \eta_1(n_1)
 &:= \eta\left(\frac{1}{2} - \frac{n_1}{2(N-1)}\right) \le 1
\label{loc6c} \\
 \eta_2(n_1)
 &:= \eta\left(\frac{1}{2} + \frac{N-n_1}{2(N-1)}\right) \le \eta_2<1.
\label{loc6d} 
\end{align}
Notice that $\eta_2$ is automatically $<1$, while $\eta_1(n_1)<1$ if 
$\lim_N  n_1/N \ne 0$.
If so, the upper bound
\begin{equation}
 E(n) \le N\Bigl( 2^{-(N-1)(1-\eta_1(n_1))} \,+\, 2^{-(N-1)(1-\eta_2)} \Bigr)
\label{loc7a}
\end{equation}
implies $N\bigl| \< C_N, W_N(\bi n)\, C_N \> \bigr|^2\longmapsto 0$
exponentially with $N\to\infty$.

The condition for which $\eta_1(n_1)<1$ is fulfilled when 
$|x_1-y_1|>\delta$; in fact, $\bi n=[N\bi y]-[N\bi x]$ and $\lim_N ([Nx_1]-[Ny_1])/N = x_1-y_1$.
On the other hand, if
$x_1=y_1$ and $n_2 = [Nx_2]-[Ny_2] \ne 0$, one explicitly computes
\begin{equation}
 N \bigl| \< C_N, W_N((0,n_2))\, C_N \> \bigr|^2 =
 N\Bigl( \cos^2\bigl(\frac{\pi n_2}{N}\bigr) \Bigr)^{N-1}.
\label{loc8}
\end{equation}
Again, the above expression goes exponentially fast to zero, if
$\lim_N n_2/N \ne 0$ which is the case if $x_2 \ne y_2$.

\section{Quantum and classical time evolutions}
\label{s5}

One of the main issues in the semi-classical analysis is to compare
if and how the quantum and classical time evolutions mimic each other
when a quantization parameter goes to zero.

In the case of classically chaotic quantum systems, the situation is
strikingly different from the case of classically integrable quantum
systems. In the former case, classical and quantum mechanics agree on
the level of coherent states only over times which scale as
$-\!\log\hbar$. As before, let $T$ denote the evolution on the
classical phase space $\c X$ and $U_T$ the unitary single step
evolution on $\Cx^N$. We formally impose the relation between the
classical and quantum evolution on the level of coherent states
through:

\begin{condition}
\label{dynloc}
\co{Dynamical localization:} 
 There exists an $\alpha>0$ such that for all choices of
 $\varepsilon>0$ and $d_0>0$ there exists an $N_0\in\Nl$ with the
 following property: if  $N > N_0$ and $k\le \alpha \log N$, then $N
 |\< U_T^kC_N(x), C_N(y)\>|^2 \le \varepsilon$ whenever $d(T^kx,y)
 \ge d_0$.
\end{condition} 

\medskip

\noindent
{\bf Remark}
\quad
The condition of dynamical localization is what is expected of
a good choice of coherent states, namely, on a time scale
logarithmic in the inverse of the
semi-classical parameter, evolving coherent states should stay localized around the
classical trajectories.
Informally, when $N\to\infty$, the quantities
\begin{equation}
\label{dyn-loc}
K_{k}(x,y) := \< U_T^kC_N(x), C_N(y)\>
\end{equation}
should behave as if $N|K_k(x,y)|^2\simeq\delta(T^kx-y)$.
The constraint $k\le \alpha\log N$ is typical of hyperbolic classical behaviour 
and comes heuristically as follows.
The maximal localization of coherent states cannot exceed the minimal
coarse-graining dictated by $1/N$; if, while evolving,
coherent states stayed localized forever around the classical trajectories, 
they would get more and more localized along the contracting direction.
Since for hyperbolic systems the increase of localization is 
exponential with Lyapounov exponent $\lambda_{Lyap}>0$, this sets 
the upper bound and indicates that $\alpha\simeq1/\lambda_{Lyap}$.
\medskip

\begin{proposition}
\label{prop3}
 Let $(\c M_N ,\Theta_N,\tau_N)$ be a general quantum
 dynamical system as defined in Section~\ref{s3} and suppose
 that it satisfies Condition~\ref{dynloc}. Let $\|X\|_2 := \sqrt{
 \tau_N(X^*X)}$, $X\in \c M_N$ denote the normalized
 Hilbert-Schmidt norm. In the ensuing topology 
 \begin{equation}
  \lim_{\substack{k,\ N\to\infty \\k< \alpha\log N}} \| \Theta_N^k
  \circ \gamma_{N \infty}(f) - \gamma_{N \infty} \circ \Theta^k(f)
  \|_2 = 0.\label{added}
 \end{equation}
\end{proposition}

\begin{proof}
One computes
\begin{align}
 &\|\Theta_N^k \circ \gamma_{N\infty}(f) - \gamma_{N\infty} \circ
 \Theta^k(f)\|_2^2 
\nonumber \\
 &= 2 N \int_{\c X} \mu(dx)\, \int_{\c X} \mu(dy)\, \overline{f(x)}\, f(y)\, 
 |\< C_N(x), C_N(y) \>|^2 
\nonumber \\
 &\quad - 2 N \Re\g e\left[\int_{\c X} \mu(dx)\, \int_{\c X}
 \mu(dy)\,  \overline{f(y)}\, f(T^k x)  |\< U_T^k C_N(x), C_N(y) \>
 |^2 \right].
\end{align}

The double integral in the first term goes to $\int \mu(dx) |f(x)|^2$.
So, we need to show that the second integral, which we shall denote by
$I_N(k)$, does the same. We will concentrate on the case of
continuous $f$, the extension to essentially bounded $f$ is straightforward.
Explicitly, selecting a ball $B(T^k x,d_0)$, one derives
\begin{align*}
 &\left| I_N(k) -
 \int_{\c X} \mu(dy)\, |f(y)|^2 \right| \\
 &=  \left| \int_{\c X} \mu(dx)\, \int \mu(dy)\, \overline{f(y)}\,
 \bigl( f(T^k x) - f(y) \bigr)\, N|\< U_T^k C_N(x), C_N(y)\>|^2
 \right| \\
 & \le  \left| \int_{\c X} \mu(dx)\, \int_{B(T^k x,d_0)} \mu(dy)\,
 \overline{f(y)} \bigl(f(T^k x) - f(y)\bigr) N|\< U_T^k C_N(x),
 C_N(y)\>|^2 \right| \\
 &+ \left| \int_{\c X} \mu(dx)
 \int_{\c X\setminus B(T^k x,d_0)} \mu(dy)
 \overline{f(y)}\bigl(f(T^k x) - f(y)\bigr) N|\< U_T^k C_N(x),
 C_N(y)\> |^2 \right|.
\end{align*}
Applying the mean value theorem and approximating the integral of the
kernel as in the proof of Proposition 3.2, we get that 
$\exists c \in B(T^k x, d_0)$ such that
\begin{align*}
 &\left| I_N(k) -
 \int_{\c X} \mu(dy)\, |f(y)|^2 \right|\\
 &\le \left| \int_{\c X} \mu(dx)\, 
 \overline{f(c)}\, \bigl(f(T^k x) -
 f(c)\bigr)\, N|\< U_T^k C_N(x), C_N(y)\>|^2 \right| \\
 &+ \left| \int_{\c X} \mu(dx) \int_{\c X\setminus B(T^k x,d_0)} 
 \mu(dy)
 \overline{f(y)} \bigl(f(T^k x) - f(y)\bigr) N|\< U_T^k C_N(x),
 C_N(y)\>|^2 \right|. 
\end{align*}
By uniform continuity we can bound the first term by some arbitrary
small $\varepsilon$, provided  we choose $d_0$ small enough. Now, for
the second integral we use our localization condition~\ref{dynloc}. 
As the constraint $k \le \alpha \log N$ has to be enforced, we
have to take a joint  limit of time and size of the system with this
constraint. In that case the second integral can  also be bounded by
an arbitrarily small $\varepsilon'$, provided $N$ is large enough. 
\end{proof}

We shall not prove the dynamical localization condition~\ref{dynloc} for 
the quantum cat maps but instead provide a direct derivation of 
formula~\eqref{added} based on the simple expression~(\ref{dyncat})
of the dynamics when acting on Weyl operators.  
For this reason, we introduce the Weyl quantization: 

\begin{definition}
\label{Weyl-q}
 Let $f$ be a function in $\g L_\mu^\infty(\Ts)$, $\Ts$ denoting
 the two--dimensional torus, whose Fourier series  
 $\hat f$ has only finitely many non-zero terms. We shall denote by
 $\r{Supp}(\hat f)$ the support of $\hat f$ in $\Ir^2$. Then, in the
 Weyl quantization scheme, one associates to $f$ the $N\times N$
 matrix
 \begin{equation*} 
  W_N(f) := \sum_{\bi k\in {\r{Supp}(\hat f)}} \hat{f}(\bi k)\,
  W_N(\bi k).
 \end{equation*}
\end{definition}
Our aim is to prove: 

\begin{proposition}
\label{prop3cat}
 Let $(\c M_N ,\Theta_N,\tau_N)$ be a sequence of quantum cat maps
 tending with $N\to\infty$ to a classical cat map with Lyapounov exponent
 $\log\lambda$; then
 \begin{equation*}
  \lim_{\substack{k,\ N\to\infty \\k< \log N/(2\log\lambda)}} \|
  \Theta_N^k \circ \gamma_{N \infty}(f) - \gamma_{N \infty} \circ
  \Theta^k(f) \|_2 = 0\quad ,
  \end{equation*}
where $\|\,\cdot\,\|_2$ is the Hilbert-Schmidt norm of Proposition 5.1.
\end{proposition}

\noindent
First we prove an auxiliary result.

\begin{lemma}
\label{Lemma1}
 If $\bi n=(n_1,n_2)\in\Ir^2$ is such that $0\le n_i\le N-1$ and
 $\lim_N  \frac{n_i}{\sqrt{N-1}}=0$, then the expectation of Weyl
 operators $W_N(\bi n)$ with respect to the state $| C_N \>$ given
 in~\eqref{coh2} is such that
 \begin{equation*}
  \lim_{N\to\infty} \< C_N, W_N(\bi n)\, C_N \> = 1.
 \end{equation*}
\end{lemma}

\begin{proof} 
The idea of the proof is to use the fact that, for large $N$, the
binomial coefficients $\binom{N-1}{j}$ contribute to the binomial sum
only when $j$ stays within a neighbourhood of $(N-1)/2$ of width
$\simeq \sqrt{N}$, in which case they can be approximated by a
normalized Gaussian function. We also notice that, by expanding the
exponents in the bounds~(\ref{loc7a})  and~(\ref{loc8}), the
exponential decay fails only if $n_{1,2}$ grow with $N$ slower 
than $\sqrt{N}$, which is surely the case for fixed finite $n$, whereby
it also follows that we can disregard the second term
in the sum comprising the contributions~(\ref{loc2}). We then write
the $j$'s in the binomial coefficients as
\begin{equation*}
 j = \Bigl[ \frac{N-1}{2} \Bigr] + k = \frac{N-1}{2} + k - \alpha, \quad
 \alpha\in\{0,{\textstyle \frac{1}{2}}\},
\end{equation*}
and consider only $k = {\r O}(\sqrt N)$. Stirling's formula
\begin{equation*}
 L! = L^{L+1/2}\, \r e^{-L}\, \sqrt{2\pi}\,  \Bigl(1 + {\r
 O}(L^{-1})\Bigr),
\end{equation*}
allows us to rewrite the first term in the r.h.s.\ of~(\ref{loc2}) as 
\begin{align}
 &\exp{\left(-\frac{1}{2}\frac{n_1^2}{N-1}\right)}\
 \sum_{k=-\left[\frac{N-1}{2}\right]}^{N-1-\left[\frac{N-1}{2}\right]+n_1}
 \frac{2\,\r e^{\frac{2\pi i}{N}n_2
 \left(k+\left[\frac{N-1}{2}\right]\right)}}{\sqrt{2\pi(N-1)}}
\nonumber \\
 &\quad\times \exp{\left(-\frac{2(k-\alpha+\frac{n_1}{2})^2}{N-1}\right)}
 \Bigl(1 + \r O(N^{-1}) + \r O((k+n_1)^3\, N^{-2}) \Bigr).
\label{lemma1.1}
\end{align}
For any fixed, finite $\bi n$, both the sum and the factor in front tend to
1, the sum becoming the integral of a normalized Gaussian.
\end{proof}
\medskip

{\setlength{\parindent}{0pt}  
\textbf{Proof of Proposition~\ref{prop3cat}:}
Given $f\in\g L_\mu^\infty(\c X)$ and $\varepsilon>0$, we choose $N_0$
such that the Fourier approximation $f_\varepsilon$ of $f$ with
$\#(\r{Supp}(\hat f)) = N_0$ is such that $\|f-f_\varepsilon\|
\le\varepsilon$, where $\|\cdot\|$ denotes the usual Hilbert space norm.
Next, we estimate
\begin{align*}
 I_N(f)
 &:= \bigl\| \Theta_N^k\circ\gamma_{N\infty}(f) -
 \gamma_{\infty N}\circ\Theta^k(f) \bigr\|_2 \\
 &\le \bigl\| \Theta_N^k\circ\gamma_{N\infty}(f-f_\varepsilon)
 \bigr\|_2 + \bigl\| \gamma_{N\infty}\circ\Theta^k(f-f_\varepsilon)
 \bigr\|_2 \\
 &\quad+ \bigl\| \Theta_N^k\circ\gamma_{N\infty}(f_\varepsilon) - 
 \gamma_{N\infty}\circ\Theta^k(f_\varepsilon) \bigr\|_2 \\
 &\le 2\|f-f_\varepsilon\| + I_N(f_\varepsilon).
\end{align*}

This follows from $\Theta_N$-invariance of the norm 
$\|\,\cdot\,\|_2$, from $T$-invariance of the measure $\mu$
and from the fact that the positivity inequality for unital
completely positive maps such as $\gamma_{N\infty}$ gives:
\begin{align*}
 \bigl\| \gamma_{N\infty}(g) \bigr\|_2^2
 &= \tau_N \bigl( \gamma_{N\infty}(g)^*\gamma_{N\infty}(g)
 \bigr) \le \tau_N\Bigl( \gamma_{N\infty}(|g|^2)\Bigr) \\
 &= \int_{\Ts} d\bi x\, |g|^2(\bi x) = \|g\|^2\quad .
\end{align*}

We now use that $f_\varepsilon$ is a function with finitely 
supported Fourier transform and, inserting the Weyl quantization of
$f_\varepsilon$, we estimate 
\begin{equation}
 I_N(f_\varepsilon) \le \bigl\| \gamma_{N\infty}(f_\varepsilon)-
 W_N(f_\varepsilon) \bigr\|_2 + \bigl\| 
 \gamma_{N\infty} \circ \Theta^k(f_\varepsilon) - 
 \Theta^k_N(W_N(f_\varepsilon)) \bigr\|_2.
\label{prop3.1}
\end{equation}
Then, we concentrate on the square of the
second term, which we denote by $G_{N,k}(f_\varepsilon)$ and 
explicitly reads
\begin{align}
 G_{N,k}(f_\varepsilon)
 &= \tau_N \bigl( \gamma_{N\infty} \circ \Theta^k(f_\varepsilon^*
 \gamma_{N\infty} \circ \Theta^k(f_\varepsilon) \bigr) + \tau_N
 \bigl( W_N(f_\varepsilon)^* W_N(f_\varepsilon)\bigr) 
\nonumber \\
 &\quad- 2\Re\g e\Bigl( \tau_N \bigl(
 \gamma_{N\infty} \circ \Theta^k(f_\varepsilon)^*
 \Theta^k_N(W_N(f_\varepsilon))\bigr) \Bigr).
\label{prop3.2}
\end{align}

The first term tends to $\|f_\varepsilon\|^2$ as $N\to\infty$,
because of Proposition~\ref{prop2} and the same is true of the second term;
indeed,
\begin{equation*}
 \tau_N \bigl( W_N(f_\varepsilon)^*W_N(f_\varepsilon) \bigr) =
 \sum_{\bi{k,q}\in\r{Supp}(\hat f_\varepsilon)} \overline{\hat
 f_\varepsilon(\bi k)}\, \hat f_\varepsilon(\bi q)\,
 {\rm e}^{\frac{i\pi}{N}\sigma(\bi q,\bi k)}\,
 \tau_N \bigl( W_N(\bi{q-k})\bigr).
\end{equation*}
Now, since $\r{Supp}(\hat f_\varepsilon)$ is finite, the vector
$\bi{k-q}$ is uniformly bounded with respect to $N$. Therefore, with
$N$ large enough, (\ref{Weyl5}) forces $\bi k=\bi q$, whence the
claim. It remains to show that the same for the third term
in~(\ref{prop3.2}) which amounts to twice the real part of
\begin{align*}
 &\int_{\Ts} d\bi x\, \overline{f_\varepsilon(A^{-k}\bi x)}
 \< C_N(\bi x), \Theta^k_N(W_N(f_\varepsilon)) C_N(\bi x)\> \\
 &\quad= \sum_{\bi p\in\c S(f_\varepsilon)}
 \overline{\hat{f}_\varepsilon(\bi p)} \< C_N, W_N(A^k\bi p) C_N\>
 \int_{\Ts} d\bi x \overline{f_\varepsilon(A^{-k}\bi x)}
 \exp\Bigl(\frac{2\pi i}{N}\sigma(A^k\bi p,[N\bi x]) \Bigr).
\end{align*}
According to Lemma~\ref{Lemma1}, the matrix element 
$\< C_N,W_N(A^k\bi p) C_N\>$ tends to 1 as $N\to\infty$ whenever
the vectorial components $(A^k\bi p)_j$, $j=1,2$, satisfy
\begin{equation*}
 \lim_N \frac{(A^k\bi p)^2_j}{N} = C_{\bi u}(\bi p) ({\bi u})_j \lim_N
 \frac{\lambda^{2k}}{N} = 0,
\end{equation*}
where we expanded $\bi p = C_{\bi u}(\bi p){\bi u} +  C_{\bi v}(\bi p)
{\bi v}$ along the stretching and squeezing eigendirections of $A$
(see Definition~\ref{ccat-map}).  This fact sets the logarithmic time
scale $k < \frac{1}{2}\frac{\log N}{\log\lambda}$.  Notice that, when
$k=0$, $G_{N,k}(f_\varepsilon)$ equals the first term
in~(\ref{prop3.1}) and this concludes the proof.  \hfill
$\blacksquare$ \par}
\medskip
\noindent
{\bf Remark} \quad The previous result essentially points to the fact
that the time evolution and the classical limit do commute over time
scales that are logarithmic in the semi--classical parameter $N$. The
upper bound of this time, which goes like
$\text{const.}\times\frac{\log N}{\log\lambda}$, is typical of quantum
chaos and is known as logarithmic {\it breaking--time}. Such a scaling
has been found numerically in~\cite{Ben04:1} 
also for discrete classical cat maps, converging in a suitable
classical limit to continuous cat maps.

\medskip
\section{Dynamical entropies}
\label{s6}

Intuitively, one expects the instability proper to the presence of
a positive Lyapounov exponent to correspond to some degree of
unpredictability of the dynamics: classically, the metric entropy of  
Kolmogorov provides the link~\cite{For92:1}.

In the usual setting, one considers partitions  $\c C = \{C_0, C_1,
\ldots, C_{q-1}\}$ of the phase space $\c X$ into  finitely many
measurable disjoint subsets $C_j$ (atoms). Under the dynamics $T$,
$\c C$ evolves into another finite partition $T(\c C) := \{
T^{-1}(C_0), T^{-1}(C_1), \ldots, T^{-1}(C_{q-1}) \}$. Moreover, by
intersecting atoms of partitions at different times one gets disjoint
atoms
\begin{equation*}
 C_{\bi i} := \bigcap_{j=0}^{k-1} T^j(C_{i_j})\quad \text{for }
 \bi i = (i_0,i_1, \ldots, i_{k-1}), 
\end{equation*}
which constitute the refined partition 
\begin{equation*}
 \c C^{(k)} := \bigvee_{j=0}^{k-1} T^j(\c C)\ .
\end{equation*}

Given the invariant measure $\mu$ on $\c X$,
the probability for the system to belong to the
atoms  $C_{i_0}$, $C_{i_1}$, \dots, $C_{i_{k-1}}$ at the successive
times  $0\le j\leq k-1$ is $\mu(C_{\bi i})$. 

In terms of symbolic dynamics, one gets a stationary stochastic
process. This amounts to a right-shift along a classical spin
half-chain with respect to a translation-invariant state. At each
site of the half-chain, one has the state space $\{0,1,\ldots,q-1\}$.
The atom $C_{\bi i}$ of the refined partition $\c C^{(k)}$ is
identified with the local configurations $\bi
i\in\{0,1,\ldots,q-1\}^k$ and has a weight 
\begin{equation*}
 \mu_{(k)}^{\c C}(\bi i) := \mu(C_{\bi i}).
\end{equation*}  
The local states $\mu_{(k)}^{\c C}$ are compatible and define a
global state on the set of extended configurations
$\{0,1,\ldots,q-1\}^\Nl$.
Such a state is invariant under
the right-shift and has a well-defined mean entropy 
\begin{equation}
 h_\mu^{\r{KS}}(T,\c C) := \lim_{k\to\infty} \frac{1}{k}
 S\bigl( \mu^{\c C}_{(k)} \bigr),
\label{17}
\end{equation}
where, for a discrete measure $\lambda$, $S(\lambda) := -\sum_j
\lambda_j\log\lambda_j$. The entropy density~(\ref{17}) is also
interpretable as average entropy  production. It consistently
measures how predictable the dynamics is on  the coarse grained 
scale provided by the finite partition $\c C$. Then,
removal of the dependence on finite partitions leads to

\begin{definition}
 The \co{KS}-entropy of a classical dynamical system $(\c X,T,\mu)$ 
 is
 \begin{equation*} 
  h^{\r{KS}}_\mu(T) := \sup_{\c C} h_\mu^{\r{KS}}(T,\c C).
 \end{equation*} 
\end{definition}

For the automorphisms of the 2-torus, we have the well-known
result~\cite{Kat99:1}:

\begin{proposition} 
\label{prop4.1}
 Let $(\g M_\mu,\Theta,\omega_\mu)$ be as in Definition~\ref{ccat-map},
 then $h^{\r{KS}}_\mu(T) = \log\lambda$.
\end{proposition}

The idea behind the notion of dynamical entropy is that information
can be obtained by repeatedly observing a system in the course of its
time evolution. Due to the uncertainty principle, or, in other words,
to non-commutativity, if observations are intended to gather
information about the intrinsic dynamical properties of quantum
systems, then  non-commutative extensions of the \co{KS}-entropy
ought first to decide whether quantum disturbances produced by
observations have to be taken into account or not.

Concretely, let us consider a quantum system described by a density
matrix $\rho$ acting on a Hilbert space $\g H$.  Via the wave packet
reduction postulate, generic measurement processes may reasonably
well be described by finite sets $\c Y = \{y_0, y_1,\ldots, y_{q-1}\}$
of bounded  operators $y_j\in \g B(\g H)$ such that $\sum_j y_j^* y_j
= \idty$. These sets are called \co{partitions of unity} and describe
the change in the state of the system caused by the corresponding
measurement process:
\begin{equation}
\label{18}
 \rho \mapsto \Gamma^*_{\c Y}(\rho) := \sum_j y_j\, \rho\, y^*_j.
\end{equation}
It looks rather natural to rely on partitions of unity to describe
the process of collecting information through repeated observations
of an evolving  quantum system~\cite{Ali94:1}. Yet, most of these
measurements interfere with the quantum evolution, possibly acting as
a source of unwanted extrinsic randomness. Nevertheless, the effect
is typically quantal and rarely avoidable.  Quite interestingly, as
we shall see later, pursuing these ideas leads to quantum
stochastic processes with a quantum dynamical entropy of their own, 
the \co{ALF}-entropy, that is also useful in a classical context.

An alternative approach~\cite{Con87:1} leads to the \co{CNT}-entropy.
This approach lacks the operational appeal of the
\co{ALF}-construction,  but is intimately  connected with the
intrinsic relaxation properties of quantum  systems~\cite{Con87:1,Nar92:1}
and possibly useful in the rapidly growing field of quantum
communication. The \co{CNT}-entropy is based on decomposing quantum
states rather than on reducing them as in~(\ref{18}).  Explicitly, if
the state $\rho$ is not a one dimensional projection, any partition
of unity $\c Y$ yields a decomposition
\begin{equation}
\label{19}
 \rho = \sum_j \tr \bigl(\rho\, y^*_jy_j\bigr)\, \frac{\sqrt\rho\,
 y^*_jy_j\sqrt\rho} {\tr\bigl(\rho\, y^*_jy_j\bigr)}.
\end{equation}
When $\Gamma^*_{\c Y}(\rho) = \rho$, reductions also provide decompositions, 
but not in general.

\subsection{\co{CNT}-entropy}

The \co{CNT}-entropy is based on decomposing quantum states into
convex linear combinations of other states.  The information content
attached to the quantum dynamics is not based on modifications of the
quantum state or on perturbations of the time evolution. Let $(\g
M,\Theta,\omega)$ represent a quantum dynamical system in  the
algebraic setting and assume $\omega$ to be decomposable.  The
construction runs as follows.

\begin{itemize}
\item
 Classical partitions are replaced by finite dimensional
 C*-algebras     $\g N$ with identity embedded into $\g M$ by
 completely positive, unity preserving (\co{cpu}) maps $\gamma: \g N
 \mapsto \g M$. Given $\gamma$, consider the cpu~maps  $\gamma_\ell
 := \Theta^\ell\circ\gamma$ that result from successive iterations of
 the dynamical automorphism $\Theta$, and associate to each of them an
 index set $I_\ell$. These index sets $I_\ell$ will be coupled to the
 cpu~maps $\gamma_\ell$ through the variational problem~\eqref{22}.
\item
 If $0\leq \ell < k$ then consider multi-indices $\bi i =
 (i_0,i_1,\ldots,i_{k-1}) \in I^{(k)} := 
 I_0\times \cdots\times I_{k-1}$ as labels of states $\omega_{\bi i}$ on
 $\g M$ and of weights $0<\mu_{\bi i}<1$  such that $\sum_{\bi i}
 \mu_{\bi i} = 1$ and $\omega = \sum_{\bi i} \mu_{\bi i}\,
 \omega_{\bi i}$. 
 These states are given by elements $0\leq x_{\bi
 i}^{\prime}\in {\g M}^{\prime}$, the commutant of $\g M$, such that
 $\sum_{\bi i} x_{\bi i}^{\prime} =  \idty_N$. Explicitly 
\begin{equation}
y\in {\g M} \longmapsto \omega_{\bi i} (y) := \frac{\omega (x_{\bi
 i}^{\prime} \, y)}{\omega (x_{\bi
 i}^{\prime})} \ , \quad \mu_{\bi i} := \omega (x_{\bi
 i}^{\prime}) \ \cdot \label{decomposition}
\end{equation}
The decomposition has be done with elements $x^{\prime}$ in the
commutant in order to ensure the positivity of the expectations $
\omega_{\bi i}$ .
\item
 From $\omega = \sum_{\bi i} \mu_{\bi i}\, \omega_{\bi i}$, one
 obtains  subdecompositions $\omega = \sum_{i_\ell\in I_\ell}
 \mu_{i_\ell}^\ell\, \omega^\ell_{i_\ell}$,  where
 \begin{equation}
  \omega^\ell_{i_\ell} := \sum_{\substack{\bi i \\ i_\ell\text{
  fixed}}} \frac{\mu_{\bi i}}{\mu^\ell_{i_\ell}}\, \omega_{\bi i}
  \qquad\text{and}\qquad
  \mu_{i_\ell}^\ell := \sum_{\substack{\bi i \\ i_\ell\text{ fixed}}}
  \mu_{\bi i}.
 \label{20}
 \end{equation}
\item 
 Since $\g N$ is finite dimensional, the states $\omega \circ
 \Theta^\ell \circ  \gamma = \omega \circ \gamma$ and 
 $\omega^\ell_{i_\ell} \circ \Theta^\ell \circ \gamma$, have finite 
 von~Neumann entropies $S(\omega\circ \gamma)$ and  
 $S(\omega^\ell_{i_\ell} \circ \Theta^\ell \circ \gamma)$. With
 $\eta(x) := -x\ \log x$ if $0<x\le1$ and $\eta(0)=0$, one defines
 the $k$~subalgebra functional
 \begin{align}
  &H_\omega(\gamma_0, \gamma_1, \ldots, \gamma_{k-1}) 
  := \sup_{\omega = \sum_{\bi i} \mu_{\bi i}\, \omega_{\bi i}}\left\{
  \sum_{\bi i} \eta(\mu_{\bi i}) - \sum_{\ell=0}^{k-1} 
  \sum_{i_\ell\in I_\ell} \eta(\mu^\ell_{i_\ell}) \right.
 \nonumber \\
 &\quad+ \left.\sum_{\ell=0}^{k-1} \Bigl(S(\omega\circ \gamma_\ell)
   - \sum_{i_\ell\in I_\ell} \mu^\ell_{i_\ell}\,
    S(\omega^\ell_{i_\ell} \circ \gamma_\ell)\Bigr)\right\}.
 \label{22}
 \end{align}
\end{itemize}

We list a number of properties of $k$-subalgebra functionals,
see~\cite{Con87:1}, that will be used in the sequel:
\begin{itemize}
\item
 \co{positivity:} $0\le H_\omega(\gamma_0, \gamma_1, \ldots,
 \gamma_{k-1})$
\item
 \co{subadditivity:} 
 \begin{eqnarray*}
 H_\omega(\gamma_0, \gamma_1, \ldots,
 \gamma_{k-1}) &\le& H_\omega(\gamma_0, \gamma_1, \ldots,
 \gamma_{\ell-1})\\
 &+& H_\omega(\gamma_\ell, \gamma_{\ell+1}, \ldots, 
 \gamma_{k-1})\ 
 \end{eqnarray*}
\item
 \co{time invariance:} $H_\omega(\gamma_0, \gamma_1, \ldots,
 \gamma_{k-1}) = H_\omega(\gamma_\ell, \gamma_{\ell+1}, \ldots,
 \gamma_{\ell+ k-1})$
\item
 \co{boundedness:} $H_\omega(\gamma_0, \gamma_1, \ldots,
 \gamma_{k-1}) \le k H_\omega(\gamma) \le k S(\omega \circ \gamma)$ 
\item
 The $k$-subalgebra functionals are invariant under 
 interchange and repetitions of arguments:
 \begin{equation}
  H_\omega(\gamma_0, \gamma_1, \ldots, \gamma_{k-1}) =
  H_\omega(\gamma_{k-1}, \ldots, \gamma_0, \gamma_0).
 \label{25}
 \end{equation}
\item
 \co{monotonicity:} If $i_\ell: \g N_\ell \mapsto \g N$, 
 $0\le\ell\le k-1$, are cpu~maps from finite dimensional
 algebras $\g N_l$ into $\g N$, then the  maps $\tilde\gamma_\ell :=
 \gamma \circ i_\ell$ are~cpu and
 \begin{equation}
  H_\omega(\tilde\gamma_0, \Theta\circ\tilde\gamma_1, \ldots, \Theta^{k-1}
  \circ \tilde\gamma_{k-1}) \le H_\omega(\gamma_0, \gamma_1, \ldots, 
  \gamma_{k-1}).
 \label{26}
 \end{equation}
\item 
 \co{continuity:} Let us consider for $\ell=0,1,\ldots,k-1$ a set of
 cpu~maps $\tilde\gamma_\ell: \g N\mapsto\g M$ such that 
 $\|\gamma_\ell - \tilde\gamma_\ell\|_\omega \le \epsilon$ for all
 $\ell$, where 
 \begin{equation}
  \|\gamma_\ell - \tilde{\gamma}_\ell\|_\omega := 
  \sup_{x\in\g N,\ \|x\|\le1}
  \sqrt{\omega\Bigl((\gamma_\ell(x) - \tilde\gamma_\ell(x))^*
  (\gamma_\ell(x) - \tilde\gamma_\ell(x))\Bigr)}\ .
 \label{27}
 \end{equation}   
 Then~\cite{Con87:1}, there exists $\delta(\epsilon)>0$ depending on the 
 dimension of the finite dimensional algebra $\c N$ and vanishing when 
 $\epsilon\to0$, such that  
 \begin{equation}
  \Bigl| H_\omega(\gamma_0, \gamma_1\ldots, \gamma_{k-1}) -
  H_\omega(\tilde{\gamma}_0, \tilde{\gamma}_1\ldots,
  \tilde{\gamma}_{k-1}) \Bigr| \le k\, \delta(\epsilon).
 \label{28}
 \end{equation}   
\end{itemize}

On the basis of these properties, one proves
the existence of the limit 
\begin{equation}
\label{new}
h^{\r{CNT}}_\omega(\theta,\gamma) :=
 \lim_k \frac{1}{k} H_\omega(\gamma_0, \gamma_1,\ldots, \gamma_{k-1})
\end{equation}
and defines~\cite{Con87:1}:

\begin{definition}
The \co{CNT}-entropy of a quantum dynamical system 
$(\g M,\Theta,\omega)$ is
\begin{equation*}
 h_\omega^{\r{CNT}}(\Theta) := \sup_\gamma
 h^{\r{CNT}}_\omega(\Theta,\gamma)\ .
\end{equation*}
\end{definition}

\subsection{\co{ALF}-entropy}

The idea underlying the \co{ALF}-entropy is that the evolution of a
quantum dynamical system can be modelled by repeated  measurements at
successive equally spaced times, the measurements  corresponding to
partitions of unity as defined in Section \ref{s6} 
which we shall refer to as {\it p.u.}, for the sake of shortness.

Such a construction
associates a quantum dynamical system  with a symbolic dynamics
corresponding  to the right-shift along a quantum spin
half-chain~\cite{Tuy97:1}.

Generic {\it p.u} $\c Y = \{y_0, y_1,\ldots, y_{\ell-1}\}$ need
not preserve the state, but
disturbances are kept under control by suitably selecting the $y_j$.
The construction of the \co{ALF}-entropy for a quantum dynamical
system $(\g M,\Theta,\omega)$ can be resumed as follows:

\begin{itemize}
\item
 One selects a sub-algebra $\g M_0 \subseteq \g M$ which is invariant under
 $\Theta$ and a
 {\it p.u.} $\c Y = \{ y_0, y_1,\ldots, y_{\ell-1} \}$ of finite
 size $\ell$ with $y_j\in\g M_0$. After $j$ time steps $\c Y$ will have 
 evolved into another {\it p.u.}
 from $\g M_0$: $\Theta^j(\c Y) := \{ \Theta^j(y_0), \Theta^j(y_1),\ldots,
 \Theta^j(y_{\ell-1}) \} \subset \g M_0$.
\item
 Every {\it p.u.} $\c Y$ of size $\ell$ gives rise to an
 $\ell$-dimensional density matrix 
 \begin{equation}
  \rho[\c Y]_{i,j} := \omega(y^*_jy_i),
 \label{29}
 \end{equation}
 with von~Neumann entropy $H_\omega[\c Y] := S(\rho[\c Y])$.
\item
 Given two {\it p.u.}  $\c Y = \{ y_0, y_1,\ldots,
 y_{\ell-1} \}$ and  $\c Z = \{ z_0, z_1,\ldots, z_{k-1}
 \}$, of sizes $\ell$ and $k$, their \co{ordered refinement} is
 the  size $\ell k$ {\it p.u.} 
 \begin{equation}
 \c Y\circ\c Z := \{ y_0z_0, y_0z_1,\ldots, y_0z_{k-1},\ldots,y_{\ell-1}
 z_{k-1} \}.
 \label{30}
 \end{equation}
\item
 Given a size~$\ell$ {\it p.u.} $\c Y$ and the \co{ordered
 time refinements}
 \begin{equation}
 \c Y^{(k)} := \Theta^{k-1}(\c Y) \circ \Theta^{k-2}(\c Y)
 \circ\cdots\circ \c Y,
 \label{31}
 \end{equation}
 the density matrices $\rho^{(k)}_{\c Y} := \rho[\c Y^{(k)}]$ define
 states on the $k$-fold tensor product $\c M_\ell^{\otimes k}$ of
 $\ell$-dimensional matrix algebras $\c M_\ell$. 
\item
 Given a {\it p.u.} $\c Y$ of size $\ell$, let 
 $\Phi_{\c Y}: \c M_\ell\otimes\g M \mapsto \g M$ and $e_M: \g
 M\mapsto\g M$, with $M\in\c M_\ell$, be linear maps defined by  
 \begin{equation}
  \Phi_{\c Y}(M\otimes x) := \sum_{i,j} y_i^*x\,y_j\, M_{ij}
  \qquad\text{and}\qquad
  e_M(x) := \sum_{i,j} y^*_i\Theta(x)\,y_j\, M_{ij}.
 \label{32}
 \end{equation}
 $\Phi_{\c Y}$ is a cpu~map, while $e_\idty(\idty)=\idty$.  
 One readily computes 
 \begin{equation*}
  \omega\Bigl( e_{M_0} \circ e_{M_1}\cdots \circ
  e_{M_{k-1}}(\idty) \Bigr) = \tr\Bigl( \rho^{(n)}_{\c Y}\, M_0\otimes 
  M_1\cdots\otimes M_{k-1} \Bigr).
 \end{equation*} 
\end{itemize} 

The states $\rho_{\c Y}^{(k)}$ are compatible and define therefore a
global  state $\omega_{\c Y}$ on the quantum spin half-chain $\c
M_\ell^\Nl$, which  is the uniform closure of $\bigcup_{n\in\Nl} \c
M_\ell^{\otimes n}$. 
Along the same line as in Section~\ref{s6}, one
associates  with the quantum dynamical system $(\g M,\Theta,\omega)$
the right shift  $\sigma$ along the quantum spin half-chain. However,
non-commutativity shows up in that $\omega_{\c Y}$ is shift-invariant
only if $\omega\Bigl( \sum_{j=0}^\ell y_j^* x y_j \Bigr) =
\omega(x)$  for all $x\in\c M_\ell^\Nl$. Note that this is the case
when {\it p.u.} give rise to decompositions of $\omega$
as in \co{CNT}-construction, (compare~(\ref{18}) and (\ref{19})). 
This leads to

\begin{definition}
 The \co{ALF}-entropy of a quantum dynamical system $(\g M,\Theta,\omega)$ 
 is 
 \begin{equation}
  h_{(\omega,\g M_0)}^{\r{ALF}}(\Theta) := \sup_{\c Y\subset\g M_0} 
  h^{\r{ALF}}_\omega(\Theta,\c Y)
  \quad\hbox{with}\quad 
  h^{\r{ALF}}_\omega(\Theta,\c Y) := \limsup_k \frac{1}{k}
  H_\omega[\c Y^{(k)}].
 \label{33}
 \end{equation}
\end{definition}

\subsection{Quantum Dynamical Entropies Compared}

In this section we outline some of the main features of both quantum
dynamical entropies. The first thing to notice is that the \co{CNT}-
and the \co{ALF}-entropy  coincide with the \co{KS}-entropy when 
$\g M =\g M_\mu$ is the  Abelian von~Neumann algebra $\g
L^\infty_\mu(\c X)$ and $(\g M,\Theta,\omega)$  represents a
classical dynamical system. The next
observation is that when, as for the quantized hyperbolic 
automorphisms of the torus considered in this paper, $\g M$ is a
finite-dimensional algebra, both the \co{CNT}- and the
\co{ALF}-entropy are zero, see~\cite{Con87:1,Ali94:1}. Consequently, if we
decide to take the strict positivity of  quantum  dynamical entropies
as a signature of quantum chaos, quantized hyperbolic 
automorphisms of the torus cannot be called chaotic.

The complete proofs of the above facts can be found
in~\cite{Con87:1} for the \co{CNT} and~\cite{Ali96:1,Ali94:1} for the
\co{ALF}-entropy. Here, we just sketch them, emphasizing those parts
that are important to the study of their classical limit.
 
\begin{proposition}
 Let $(\g M_\mu,\Theta,\omega_\mu)$ represent a classical dynamical
 system. Then, with the notations of the previous sections
 \begin{equation*}   
  h^{\r{CNT}}_{\omega_\mu}(\Theta) = h^{\r{KS}}_\mu(T) =
  h^{\r{ALF}}_{(\omega_\mu, \g M_\mu)}(\Theta).
 \end{equation*}
\end{proposition}

\begin{proof}
\smallskip
 
\co{CNT-Entropy.}
In this case, $h^{\r{CNT}}_{\omega_\mu}(\Theta)$ is computable by using
natural embeddings  of  finite dimensional subalgebras of $\g M_\mu$
rather than generic cpu~maps  $\gamma$. Partitions $\c C
= \{ C_0, C_1,\ldots, C_{n-1} \}$ of $\c X$ can be
identified with the finite dimensional subalgebras $\c N_{\c C}\in\g
M_\mu$ generated by the  characteristic functions $\chi_{C_j}$ of the
atoms of the partition, with $\omega_\mu(\chi_C) = \mu(C)$. 
Also, the refinements $\c C^{(k)}$ of the evolving partitions $T^{-j}(\c
C)$  correspond to the subalgebras $\c N^{(k)}_{\c C}$ generated by 
$\chi_{C_{\bi i}} = \prod_{j=0}^{k-1} \chi_{T^{-j}(C_{i_j})}$.

Thus, if $\imath_{\c N_{\c C}}$ embeds $\c N_{\c C}$ into $\g M_\mu$,
then $\omega_\mu \circ \imath_{\c N_{\c C}}$ corresponds to
the state $\omega_\mu \restriction \c N_{\c C}$, which is obtained 
by restriction of $\omega_\mu$ to $\c N_{\c C}$ and
is completely determined by the expectation values  
$\omega_\mu(\chi_{C_j})$, $1\le j\le n-1$. 

Further, identifying the cpu~maps
$\gamma_\ell = \Theta^\ell \circ \imath_{\c N_{\c C}}$
with the corresponding 
subalgebras $\Theta^\ell(\c N)$,
$h^{\r{CNT}}_{\omega_\mu}(\Theta) = h^{\r{KS}}_\mu(T)$ 
follows from 
\begin{equation}
 H_\omega(\c N_{\c C}, \Theta(\c N_{\c C}),\ldots, \Theta^{k-1}(\c
 N_{\c C})) =  S_\mu(\c C^{(k)}), \quad\forall\, \c C,
\label{34}
\end{equation}
see~(\ref{17}).
In order to prove~(\ref{34}), we decompose the reference state as 
\begin{equation*}
 \omega_\mu = \sum_{\bi i} \mu_{\bi i}\, \omega_{\bi i}
 \qquad\text{with}\qquad
 \omega_{\bi i}(f) := \frac{1}{\mu_{\bi i}} \int_{\c X} \mu(dx)\, 
 \chi_{C_{\bi i}}(x)\, f(x)
\end{equation*}  
where $\mu_{\bi i} = \mu(C_{\bi i})$, see~(\ref{20}). Then, 
$\sum_{\bi i} \eta(\mu_{\bi i}) = S_\mu(\c C^{(k)})$.

On the other hand, 
\begin{equation*}
 \omega^\ell_{i_\ell}(f) =
 \frac{1}{\mu^\ell_{i_\ell}} \int_{\c X} \mu(dx)\,
 \chi_{T^{-\ell}(C_{i_\ell})}(x)\, f(x)
 \qquad\text{and}\qquad
 \mu^\ell_{i_\ell} = \mu(C_{i_\ell}).
\end{equation*}
It follows that  $\omega_\mu \circ \imath_{\c
N_{\c C}} = \omega_\mu \restriction \c N_{\c C}$ is the discrete  measure
$\{ \mu^\ell_0, \mu^\ell_1,\ldots, \mu^\ell_{n-1} \}$ for all 
$\ell=0,1,\ldots,k-1$ and, finally, that  $S(\omega^\ell_{i_\ell}
\circ \gamma_\ell) = 0$ as $\omega^\ell_{i_\ell} \circ j_\ell =
\omega^\ell_{i_\ell} \restriction  \Theta^\ell(\c N_{\c C})$ is a discrete
measure with values $0$ and $1$.
\smallskip

\co{ALF-Entropy.}  One expects that~(\ref{33}), computed over all
possible {\it p.u.} from $\g M_\mu$ should equal
$h^{\r{KS}}_\mu(T)$. Notice, however, that, even if the dynamical
system is classical, still~(\ref{33})  has to be computed within the
non-commutative setting of density matrices  as in~(\ref{29}).
In~\cite{Ali96:1}, it is shown that  $h^{\r{ALF}}_{(\omega,\g
M_\mu)}(\Theta) =  h^{\r{KS}}_\mu(T)$.
\end{proof}

In the particular case of the hyperbolic automorphisms of the torus,
we may restrict our attention to {\it p.u.}
whose elements
belong to the $\ast$-algebra $\c D_\mu$ of complex functions $f$ 
on $\Ts$ such that the support of $\hat f$ is bounded:
\begin{equation*}
 h^{\r{KS}}_\mu(T) = h^{\r{ALF}}_{(\omega_\mu, \g M_\mu)}(\Theta) =
 h^{\r{ALF}}_{(\omega_\mu, \c D_\mu)}(\Theta).
\end{equation*} 
Remarkably, the computation of the classical \co{KS}-entropy via the
quantum mechanical \co{ALF}-entropy yields a proof of
Proposition~\ref{prop4.1} that is much simpler than the
standard ones~\cite{Arn68:1,Wal82:1}.

\begin{proposition}
\label{prop4.3}
 Let $(\g M,\Theta,\omega)$ be a quantum dynamical system with
 $\g M$, a finite dimensional C*-algebra, then, 
 \begin{equation*}
  h^{\r{CNT}}_\omega(\Theta) = 0
  \qquad\text{and}\qquad
  h^{\r{ALF}}_{(\omega,\g M)}(\Theta) = 0.
 \end{equation*} 
\end{proposition}

\begin{proof}

\co{CNT-Entropy:} as in the commutative case,
$h^{\r{CNT}}_\omega(\Theta)$ is computable by means of cpu~maps
$\gamma$ that are the natural embeddings $\imath_{\c N}$ of
subalgebras $\c N\subseteq\g M$ into $\g M$. Since each
$\Theta^\ell(\c N)$ is obviously contained in the algebra  $\c
N^{(k)}\subseteq\g M$  generated by the subalgebras $\Theta^j(\c N)$,
$j=0,1,\ldots,k-1$, from the properties of the $k$-subalgebra
functionals $H$ and identifying again the natural embeddings $\gamma_\ell
:= \Theta^\ell\circ \imath_{\c N}$ with the  subalgebras
$\Theta^\ell(\c N)\subseteq\g M$, we derive
\begin{align*}
  H_\omega(\c N, \Theta(\c N),\ldots, \Theta^{k-1}(\c N))
  &\le H_\omega(\c N^{(k)}, \c N^{(k)},\ldots, \c N^{(k)}) \\
  &\le H_\omega(\c N^{(k)}) \le q S\bigl(\omega \restriction \c
  N^{(k)} \bigr) \le \log d,
\end{align*}
where $\g M \subseteq \c M_d$. In fact, $\omega \restriction \c N$
amounts to a density matrix  with eigenvalues $\lambda_\ell$ and
von~Neumann entropy  $S(\omega \restriction \c N) = -\sum_{\ell=1}^d
\lambda_\ell \log\lambda_\ell \le \log d$. Therefore, for all $\c
N\subseteq\g M$, $h^{\r{CNT}}_\omega(\Theta, \c N) = 0$.
\smallskip

\co{ALF-Entropy:} Let the state $\omega$ on $\c M_d$ be given by
$\omega(x) = \tr(\rho\,x)$, where $\rho$ is a density matrix in $\c
M_d$. Given a partition of unity $\c Y$ of size $\ell$, the cpu~map
$\Phi_{\c Y}$ in~(\ref{32}) can be used to define a state  
$\Phi_{\c Y}^*(\rho)$ on $\c M_\ell\otimes\g M$ which is dual to $\omega$:
\begin{equation*} 
 \Phi^*_{\c Y}(\rho)(M\otimes x) = \tr{\Bigl( \rho\, \Phi_{\c
 Y}(M\otimes x) \Bigr)},\quad M\in\c M_\ell,\ x\in\g M.
\end{equation*}
Since $\sum_{j=0}^\ell y^*_jy_j = \idty$, it follows that  
$\Phi^*_{\c Y}(\rho^k) = \bigl( \Phi^*_{\c Y}(\rho) \bigr)^k$.
Therefore, $\rho$ and $\Phi_{\c Y}^*(\rho)$ have the same spectrum,
apart possibly from the eigenvalue zero, and thus  the same
von~Neumann entropy. Moreover, $\Phi^*_{\c Y}(\rho) \restriction \c
M_\ell = \rho[\c Y]$ and  $\Phi^*_{\c Y}(\rho) \restriction \g M =
\Gamma^*_{\c Y}(\rho)$ as in~(\ref{18}). Applying the triangle
inequality for the entropy~\cite{Ohy93:1} 
\begin{equation*} 
 S\bigl( \Phi^*_{\c Y}(\rho) \bigr) \ge  \Bigl| S\bigl( \Phi^*_{\c
 Y}(\rho) \restriction \c M_\ell \bigr) - S\bigl( \Phi^*_{\c Y}(\rho)
 \restriction \g M \bigr) \Bigr|,
\end{equation*} 
one obtains $S(\rho[\c Y]) \le 2\log d$. Finally, as evolving
{\it p.u.} $\Theta^j(\c Y)$ and their  ordered
refinements~(\ref{30}), (\ref{31}) remain in  $\g M$, one gets 
\begin{equation*}
 \limsup_k \frac{1}{k} H_\omega[\c Y^{(k)}] = 0,\quad \c
 Y\subset\g M.
\end{equation*} 
\end{proof}

From the considerations of above, it is clear that the main field of
application  of the \co{CNT}- and \co{ALF}-entropies are infinite
quantum systems, where the  differences between  the two come to the
fore~\cite{Ali95:1}. The former has been proved to be useful to connect
randomness with clustering  properties and  asymptotic commutativity.
A rather strong form of clustering and asymptotic Abeliannes is
necessary to have a non-vanishing
\co{CNT}-entropy~\cite{Nar92:1,Nar95:1,Nes00:1}.  
In particular, the infinite dimensional quantization of the
automorphisms of the torus has vanishing \co{CNT}-entropy for most of
irrational  values of the deformation parameter $\phi$, whereas,
independently of the value of $\phi$, the \co{ALF}-entropy is always
equal to the positive Lyapounov exponent. These results reflect the
different perspectives upon which the two constructions are based. 

\section{Classical limit of quantum dynamical entropies}
\label{s7}

Proposition~\ref{prop4.3} confirms the intuition that finite
dimensional, discrete time, quantum dynamical systems, however
complicated the distribution  of their quasi-energies might be,
cannot produce enough information  over large times to generate a
non-vanishing entropy per unit time. This is due to the fact that,
despite the presence of almost random features over finite intervals,
the time evolution cannot bear random signatures if watched long
enough, because almost periodicity would always prevail
asymptotically.

In this section we take the
the \co{CNT} and the \co{ALF}-entropy as
good indicators of the degree of randomness of a quantum dynamical
system. Then, we show that underlying classical chaos plus
Hilbert space finiteness make a characteristic logarithmic time scale emerge 
over which these systems can be called
chaotic.

\subsection{CNT-entropy}

\begin{theorem}
 Let $(\c X,T,\mu)$ be a classical dynamical system 
 which is the classical limit of a sequence of finite dimensional
 quantum dynamical systems $(\c M_N,\Theta_N,\tau_N)$. 
 We also assume that the dynamical localization
 condition~\ref{dynloc} holds. If
 \begin{enumerate}
 \item 
  $\c C = \{ C_0, C_1,\ldots, C_{q-1} \}$ is a finite measurable partition
  of $\c X$,
 \item 
  $\c N_{\c C} \subset \g M_\mu$ is the finite dimensional subalgebra
  generated by the characteristic functions $\chi_{C_j}$ of the atoms
  of $\c C$,
 \item 
  $\imath_{\c N_{\c C}}$ is the natural embedding of $\c N_{\c C}$
  into $\g M_\mu = \g L_\mu(\c X)$, $\gamma_{N\infty}$ the anti-Wick
  quantization map and 
  \begin{equation*}
   \gamma^\ell_{\c C} := \Theta^\ell_N \circ \gamma_{N\infty}
   \circ \imath_{\c N_{\c C}},\quad \ell=0,1,\ldots,k-1,
  \end{equation*}
 \end{enumerate}
 then there exists an $\alpha$ such that
 \begin{equation*}
  \lim_{\substack{k,\ N\to \infty \\ k\le \alpha \log N}} \frac{1}{k} 
  \left| H(\gamma^0_{\c C}, \gamma^1_{\c C},\ldots, 
  \gamma^{k-1}_{\c C}) - S_\mu\bigl( \c C^{(k)} \bigr) \right| = 0.
 \end{equation*}
\end{theorem}

\begin{proof}
We split the proof in two parts:
\begin{enumerate}
\item 
 We relate the quantal evolution $\gamma^\ell_{\c C} = \Theta_N^\ell
 \circ \gamma_{N\infty} \circ \imath_{\c N_{\c C}}$ to the
 classical evolution $\tilde\gamma_{\c C}^\ell := \gamma_{N\infty}
 \circ \Theta^\ell \circ \imath_{\c N_{\c C}}$ using the continuity
 property of the entropy functional.
\item 
 We find an upper and a lower bound to the entropy functional that 
 converge to  the \co{KS}-entropy in the long time limit.
\end{enumerate}

We define for convenience the algebra $\c N_{\c C}^\ell := 
\Theta^\ell(\c N_{\c C})$ and the algebra $\c N^{(k)}_{\c C}$
corresponding to the refinements  $\c C^{(k)} =
\bigvee^{k-1}_{\ell=0} T^{-\ell}(\c C)$ which consist of
atoms $\c C_{\bi i} := \bigcap^{k-1}_{\ell=0} T^{-\ell}(C_{i_\ell})$
labeled by the multi-indices $\bi i = (i_0, i_1,\ldots, i_{k-1})$. 
Thus the algebra $\c N^{(k)}_{\c C}$ is generated by the
characteristic functions $\chi_{C_{\bi i}}$. 
\smallskip

\textbf{Step 1}
\smallskip

The maps $\gamma^\ell_{\c C} $ and $\tilde{\gamma}^\ell_{\c C}$
connect the quantum and classical time evolution. Indeed, using 
Proposition~\ref{prop3} 
\begin{equation*}
 k \le \alpha \log N \Rightarrow \|\Theta^k_N \circ  \gamma_{N\infty}
 \circ \imath_{\c N_{\c C}}(f) -  \gamma_{N\infty} \circ \Theta^k
 \circ \imath_{\c N_{\c C}}(f) \|_2 \le \varepsilon, 
\end{equation*}
or
\begin{equation*}
 k \le \alpha \log N \Rightarrow \| \gamma^k_{\c C} - 
 \tilde\gamma^k_{\c C} \|_2 \le \varepsilon 
\end{equation*}

This in turn implies, due to strong continuity,
\begin{equation*}
 \left| H(\gamma^0_{\c C}, \gamma^1_{\c C},\ldots, \gamma^{k-1}_{\c
 C}) - H(\tilde\gamma^0_{\c C}, \tilde\gamma^1_{\c C},\ldots,
 \tilde\gamma^{k-1}_{\c C}) \right| \le k \delta(\varepsilon)
\end{equation*}
with $\delta(\varepsilon)>0$ depending on the dimension of the space
$\c N_{\c C}$ and vanishing when $\varepsilon \to 0$. From now on we
can concentrate on the classical evolution and benefit from its
properties.
\smallskip

\textbf{Step 2, upper bound}
\smallskip

We now show that 
\begin{equation*}
 H(\tilde\gamma^0_{\c C}, \tilde\gamma^1_{\c
 C},\ldots, \tilde\gamma^{k-1}_{\c C}) \le S_\mu(\c C^{(k)}).
\end{equation*}
Notice that we can embed $\c N^\ell_{\c C}$ into $\g M_\mu$ by first
embedding it into $\c N^{(k)}_{\c C}$ with  $\imath_{\c N^\ell_{\c C}
\c N^{(k)}_{\c C}}$ and then embedding $\c N^{(k)}_{\c C}$ into $\g
M_\mu$ with $\imath_{\c N^{(k)}_{\c C}}$:
\begin{equation*}
 \imath_{\c N^\ell_{\c C}} = \imath_{\c N^{(k)}_{\c C}} \circ
 \imath_{\c N^\ell_{\c C} \c N^{(k)}_{\c C}}. 
\end{equation*}
We now estimate:
\begin{align}
 &H(\tilde\gamma^0_{\c C}, \tilde\gamma^1_{\c C},\ldots, 
 \tilde\gamma^{k-1}_{\c C}) 
\nonumber \\ 
 &\quad = H(\gamma_{N\infty} \circ \imath_{\c N^{(k)}_{\c C}} 
 \circ \imath_{\c N^0_{\c C} \c N^{(k)}_{\c C}},\ldots, 
 \gamma_{N\infty} \circ \imath_{\c N^{(k)}_{\c C}}  \circ \imath_{\c
 N^{k-1}_{\c C} \c N^{(k)}_{\c C}}) 
\nonumber \\
 &\quad\le H(\gamma_{N\infty} \circ \imath_{\c N^{(k)}_{\c
 C}},\ldots, \gamma_{N\infty} \circ  \imath_{\c N^{(k)}_{\c C}}) 
\le H(\gamma_{N\infty} \circ \imath_{\c N^{(k)}_{\c
 C}}) 
\nonumber \\
 &\quad\le S\left( \tau_N \circ \gamma_{N\infty} \circ \imath_{\c
 N^{(k)}_{\c C}} \right).
\end{align}

The first inequality follows from monotonicity of the entropy functional, the
second from invariance under repetitions and the third from boundedness in terms
of von~Neumann entropies. The state $\tau_N \circ \gamma_{N\infty} 
\circ \imath_{\c N^{(k)}_{\c C}}$ takes the values
\begin{align*}
 &\tau_N \bigl( \gamma_{N\infty}(\chi_{C_{\bi i}}) \bigr) = 
 \tau_N \Bigl( N \int_{\c X} \mu(dx)\,\chi_{C_{\bi i}}(x)\,
 |C_N(x)\> \< C_N(x)| \Bigr) \\
 &\quad= \int_{\c X} \mu(dx)\, \chi_{C_{\bi i}}(x)\, \< C_N(x),
 C_N(x)\> = \omega_{\mu}(\chi_{C_{\bi i}}) = \mu(C_{\bi i}). 
\end{align*}
This gives, together with $S(\mu(C_{\bi i})) = S_\mu(\c C^{(k)})$,
the desired upper bound.
\smallskip

\textbf{Step 2, lower bound}
\smallskip

We show that $\forall \varepsilon>0$ there exists an $N$ such that
\begin{equation*}
 H(\tilde\gamma^0_{\c C}, \tilde\gamma^1_{\c C},\ldots, 
 \tilde\gamma^{k-1}_{\c C}) \ge S_\mu(\c C^{(k)}) - k \varepsilon.
\end{equation*}
As $H(\tilde\gamma^0_{\c C}, \tilde\gamma^1_{\c
C},\ldots,  \tilde\gamma^{k-1}_{\c C})$ is defined as a supremum over
decompositions of the state $\tau_N$, we can construct a lower
bound by picking a good decomposition. Consider the decomposition
$\tau_N = \sum_{\bi i} \mu_{\bi i}\, \omega_{\bi i}$ with
\begin{align*}
 &\omega_{\bi i}: \c M_N \ni x \mapsto \omega_{\bi i}(x) :=
 \frac{\tau_N (\gamma_{N\infty}(\chi_{C_{\bi i}}))(x)}
 {\tau_N (\gamma_{N\infty}(\chi_{C_{\bi i}}))} \\
 &\mu_{\bi i} := \tau_N (\gamma_{N\infty}(\chi_{C_{\bi i}}))
\end{align*}
and the subdecompositions $\tau_N = \sum_{j_\ell}\mu^\ell_{j_\ell}\, 
\omega^\ell_{j_\ell}$, $\ell=0,1,\ldots,k-1$, with
\begin{align*}
 &\omega^\ell_{j_\ell}: \c M_N \ni x \mapsto \omega^\ell_{j_\ell}(x) :=
 \frac{\tau_N (\gamma_{N\infty}(\chi_{T^{-\ell}(C_{j_\ell})}))(x)}
 {\tau_N \bigl(\gamma_{N\infty}(\chi_{C_{j_l}}) \bigr)} \\
 &\mu^\ell_{j_\ell} := \tau_N (\gamma_{N\infty}(\chi_{C_{j_\ell}})).
\end{align*} 
In comparison with~\eqref{decomposition}, it is not necessary to go to
the commutant for one can use the cyclicity property of the trace. 
We then have:
\begin{equation*}
 H(\tilde\gamma^0_{\c C}, \tilde\gamma^1_{\c C},\ldots, 
 \tilde\gamma^{k-1}_{\c C}) \ge S_\mu(\c C^{(k)}) -\sum^{k-1}_{i=0}
 \sum_{i_\ell \in I_\ell} \mu^\ell_{i_\ell}\,  S(\omega^\ell_{i_\ell}
 \circ \tilde\gamma^\ell_{\c C}).
\end{equation*}
The inequality stems from the fact that $H(\tilde\gamma^0_{\c C}, 
\tilde\gamma^1_{\c C},\ldots, 
\tilde\gamma^{k-1}_{\c C} )$ is a supremum, whereas the middle terms
in the original definition of the entropy functional drop out because
they are equal in magnitude but opposite in sign. For
$s=0,1,\ldots,k-1$, $\omega^\ell_{i_\ell} \circ \tilde\gamma^\ell_{\c
C}$ takes on the values 
\begin{equation*}
 \omega^\ell_{i_\ell}(\gamma_{N\infty}(\chi_{T^{-\ell}(C_s)})) = 
 \mu_{i_\ell}^{-1} \tau_N 
 \Bigl(\gamma_{N\infty}(\chi_{T^{-\ell}(C_{j_\ell})})\,
 \gamma_{N\infty}(\chi_{T^{-\ell}(C_s)}) \Bigr). 
\end{equation*}
Due to Proposition~\ref{prop2}, these converge to $\omega_\mu(
\chi_{T^{-\ell}(C_{j_\ell})}\, \chi_{T^{-\ell}(C_s)}) = \delta_{i_\ell,s}$. 
This means that in
the limit the von~Neumann entropy will be zero. Or stated more carefully,
$\forall \varepsilon': \exists N'$ such that 
\begin{equation*}
 \sum^{k-1}_{i=0} \sum_{i_\ell\in I_\ell} \mu^\ell_{i_\ell}\,
 S(\omega^\ell_{i_\ell} \circ \tilde\gamma^\ell_{\c C}) \le k
 \varepsilon'.
\end{equation*}
We thus obtain a lower bound.
\smallskip

Combining our results and choosing $\tilde N := \max(N,N')$, we
conclude
\begin{equation*}
 S_\mu(\c C^{(k)}) - k \varepsilon' - k \delta(\varepsilon) \le
 H(\gamma^0_{\c C}, \gamma^1_{\c C},\ldots, 
 \gamma^{k-1}_{\c C}) \le S_\mu(\c C^{(k)}) + k \delta(\varepsilon).
\end{equation*}
\end{proof}

\subsection{ALF-entropy}

\begin{theorem}
 Let $(\c X,T,\mu)$ be a classical dynamical system
 which is the classical limit of a sequence of finite dimensional
 quantum dynamical systems $(\c M_N,\Theta_N,\tau_N)$. 
 We also assume that the dynamical localization
 condition~\ref{dynloc} holds. If
 \begin{enumerate}
 \item 
  $\c C = \{ C_0, C_1,\ldots, C_{q-1} \}$ is a finite measurable
  partition of $\c X$,
 \item 
  $\c Y_N = \{ y_0, y_1,\ldots, y_q \}$ is a bistochastic partition of
  unity, which is the quantization  of the previous partition, namely
  $y_i = \gamma_{N\infty}(\chi_{C_i})$  for $i=0,1,\ldots,q-1$  and
  $y_q := \sqrt{\idty-\sum_{i=0}^{q-1} y_i^*y_i}$,
 \end{enumerate} 
 then there exists an $\alpha$ such that
 \begin{equation*}
  \lim_{\substack{k,N\to\infty \\ k\le\alpha\log N}} \frac{1}{k}
  \left| H[\c Y_N^{(k)}] - S_\mu(\c C^{(k)}) \right| = 0.
 \end{equation*}
\end{theorem}

\begin{proof}
First notice that $\c Y_N = \{ y_0, y_1,\ldots, y_q \}$ is indeed a 
bistochastic partition. We have
\begin{align*}
 &y_i^* = \gamma_{N\infty}(\chi_{C_i})^* =
 \gamma_{N\infty}(\overline{\chi_{C_i}}) =
 \gamma_{N\infty}(\chi_{C_i}) = y_i \\ 
 &0 \le \gamma_{N\infty}(\chi_{C_i})^2 = y_i^2 \le
 \gamma_{N \infty}(\chi_{C_i}^2) = \gamma_{N \infty}(\chi_{C_i})
\end{align*}
Summing the last line over $i$ from $1$ to $q-1$, we see that
$\sum_{i=1}^{q-1} y_i^2 \le \idty$, This means that $\{ y_0,
y_1,\ldots, y_{q-1} \}$ is not a partition of unity,  but we can use
this property to define an extra  element $y_q$ which completes it to
a bistochastic partition of unity,  $\c Y_N = \{ y_0, y_1,\ldots,
y_q \}$:
\begin{equation*}
 y_q := \sqrt{\idty-\sum_{i=0}^{q-1} y_i^*y_i}
\end{equation*}
The bistochasticity is a useful property because it implies
translation invariance of the state on the quantum spin chain, state which
arises during the construction of the \co{ALF}-entropy.

The density matrix $\rho[\c Y^{(k)}]$ of the refined partition reads
\begin{align*}
 \rho\left[ \c Y^{(k)} \right] 
 &= \sum_{\bi{i,j}} \rho\left[ \c Y^{(k)} \right]_{\bi{i,j}} |e_{\bi
 i}\> \< e_{\bi j}| \\
 &= \sum_{\bi{i,j}} \tau_N \Bigl( y_{j_1}^* \cdots
 \Theta_N^{k-1}(y_{j_k}^*) \Theta_N^{k-1}(y_{i_k}) \cdots y_{i_1} \Bigr) 
 |e_{\bi i}\> \< e_{\bi j}|
\end{align*}

Now we will expand this formula using the operators $y_i$ defined above, 
the quantities $K_a(x,y)$ defined in~(\ref{dyn-loc})
and controlling  the element $y_q$ as follows:
\begin{align}
 \|y_q\|_2^2 
 &= \Bigl\| \sqrt{\idty - \sum_{i=1}^{q-1} y_i^*y_i} \Bigr\|_2^2 =
 \tau_N \Bigl( \idty - \sum_{i=1}^{q-1} y_i^*y_i \Bigr) 
\nonumber\\
 &= \int dy dz\, \sum_{i\ne j} \chi_{i}(y)\, \chi_{j}(z)\, N
 |K_0(y,z)|^2.
\label{100}  
\end{align}
Thus, in the limit of large $N$, $N |K_0(y,z)|^2$ is just $\delta(y-z)$ 
(see~(\ref{dyn-loc})) so 
that~(\ref{100}) tends to $\int dz\, \sum_{i\ne j} \chi_{i}(z)\,
\chi_{j}(z) = 0$ and we can  consistently
neglect those entries of $\rho[{\c Y}^{(k)}]$ containing
$y_q$.

By means of the properties of coherent states, 
we write out explicitly the elements of the density matrix
\begin{align}
\nonumber
 &\rho\left[\c Y^{(k)}\right]_{\bi{i,j}} = N^{2k-1}\int d\bi y d\bi
 z\, \prod_{\ell=1}^k \chi_{C_{j_\ell}}(y_\ell)
 \chi_{C_{i_\ell}}(z_\ell)\ \times\\
 &\quad
 \times K_0(z_1,y_1)\,
 \left(\prod_{p=1}^{k-1} K_1(y_p,y_{p+1})\right)\, 
 K_0(y_{k},z_{k})\,
 \left(\prod_{q=1}^{q-1} K_{-1}(z_{k-q+1},z_{k-q})\right)\quad .
 \label{matr-el}
\end{align}
We now use that for $N$ large enough,  
\begin{equation}
\label{tel} 
 \Bigl| N\int dy\,
 \chi_{C}(y)\, K_m(x,y)\, K_n(y,z) -
 \chi_{T^{-m}C}(x)\, K_{m+n}(x,z) \Bigr| \le \varepsilon_m(N)
 \, ,
\end{equation}
where $\varepsilon_m(N)\to 0$ with $N\to\infty$ uniformly in
$x,y\in\c X$.
This is a consequence of the dynamical localization condition~\ref{dynloc}
and can be rigorously proven in the same way  as Proposition~\ref{prop1}.
However, the rough idea is the following: from the property 3.1.3 of coherent
states, one derives
\begin{eqnarray*}
&&
N\int dy\, \chi_{C}(y)\, K_m(x,y)\, K_n(y,z)=K_{m+n}(x,z)\\
 &&\qquad
 +N\int dy\, \Bigl(
 \chi_{C}(y)-1\Bigr)\, K_m(x,y)\, K_n(y,z)\ .
\end{eqnarray*}
For large $N$, the
condition~\ref{dynloc} makes the integral in~(\ref{tel}) negligible small
unless $x\in T^{-m}(C)$, in which case it 
is the second integral in the formula of
above which can be neglected.

By applying ~(\ref{tel}) to the couples of products in~(\ref{matr-el})
one after the other, we finally arrive at the upper bound
\begin{equation*} 
 \bigl| \rho\left[\c Y^{(k)}\right]_{\bi{i,j}} - \delta_{\bi{i,j}}\, 
 \mu(C_{\bi i}) \bigr| \le \Bigl( 2 \sum_{m=1}^{k} \varepsilon_m(N) + 
 \varepsilon_0(N) \Bigr) =: \epsilon(N),  
\end{equation*}
where $C_{\bi i} := \bigcap_{\ell=1}^k T^{-\ell+1}C_{i_\ell}$ is an element 
of the partition $\c C^{(k)}$. 

We now set $\sigma\left[\c C^{(k)}\right] := \sum_{\bi i} \mu(C_{\bi
i}) |e_{\bi i}\>  \< e_{\bi i}|$ and use the following estimate: let
$A$ be an arbitrary matrix of dimension $d$ and let
$\{e_1, e_2,\ldots, e_d\}$ and $\{f_1, f_2,\ldots, f_d\}$ be two
orthonormal bases of $\Cx^d$, then
$\|A\|_1:=\tr |A| \le \sum_{i,j} |\< e_i,A\,f_j\>|$.
This yields 
\begin{equation*} 
\Delta(k):= \| \rho\left[\c Y^{(k)}\right] - \sigma\left[\c C^{(k)}\right] \|_1 
 = \tr \Bigl| \rho\left[\c Y^{(k)}\right] - \sigma\left[\c C^{(k)}\right]
 \Bigr|  \le  q^{2k} \epsilon(N).
\end{equation*}
Finally, by the continuity of the von~Neumann entropy~\cite{Fan73:1}, we get 
\begin{align*}
 \left|S(\rho\left[\c X^{(k)}\right]) - S(\sigma\left[\c C^{(k)}\right])\right|
 \leq \Delta(k)\log q^k\,+\,\eta(\Delta(k))\ .
\end{align*}
Since, from $k \le \alpha \log N$, $q^{2k}\leq N^{2\alpha\log q}$, if
we want the bound $q^{2k}\epsilon(N)$  to converge to zero with
$N\to\infty$, the parameter $\alpha$ has to be chosen accordingly.
Then, the result follows because the von~Neumann entropy of $\sigma$
reduces to the Shannon entropy of the refinements of the classical
partition.
\end{proof}
\section{Conclusions}
In this paper, we have shown that both the CNT and ALF entropies
reproduce the Kolmogorov-Sinai invariant if we observe a strongly
chaotic system at a very short time scale. However, due to the
discreteness of the spectrum of the quantizations, we know that
saturation phenomena will appear. It would be interesting to study
the scaling behaviour of the quantum dynamical entropies in the
intermediate region between the random breaking time and the
Heisenberg time. This will, however, require quite different
techniques than the coherent states approach.  
\bigskip 
\bigskip

\noindent
\textbf{\large Acknowledgements}
Two of the authors (F.B., V.C.) wish to express their gratitude 
to the Institute of Theoretical Physics of the University of Leuven for 
its hospitality and financial support.
\bigskip
\bigskip

\singlespacing

\end{document}